\documentclass[12pt]{iopart}

\usepackage{epsfig}
\newcommand{\pard}[2]{\frac{\partial {#1}}{\partial {#2}}}
\newcommand{\derv}[2]{\frac{d {#1}}{d {#2}}}

\begin{document}

\title{Bidirectional transport and pulsing states in a multi-lane ASEP model}

\author{Congping Lin$^{1,2}$, Gero Steinberg$^2$ and Peter Ashwin$^1$}

\address{$^{1}$ Centre for Systems, Dynamics and Control, University of Exeter, Exeter EX4 4QF, UK}

\address{$^{2}$ Department of Biosciences, University of Exeter, Exeter EX4 4QD, UK}

\eads{cl336@exeter.ac.uk, P.Ashwin@exeter.ac.uk and G.Steinberg@exeter.ac.uk}

%\date{}
%\maketitle

\begin{abstract} In this paper, we introduce an ASEP-like transport model for bidirectional motion of particles on a multi-lane lattice. The model is motivated by {\em in vivo} experiments on organelle motility along a microtubule (MT), consisting of thirteen protofilaments, where particles are propelled by molecular motors (dynein and kinesin). In the model, organelles (particles) can switch directions of motion due to ``tug-of-war'' events between counteracting motors. Collisions of particles on the same lane can be cleared by switching to adjacent protofilaments (lane changes).

We analyze transport properties of the model with no-flux boundary conditions at one end of a MT (``plus-end'' or tip). We show that the ability of lane changes can affect the transport efficiency and the particle-direction change rate obtained from experiments is close to optimal in order to achieve efficient motor and organelle transport in a living cell. In particular, we find a nonlinear scaling of the mean {\em tip size} (the number of particles accumulated at the tip) with injection rate and an associated phase transition leading to {\em pulsing states} characterized by periodic filling and emptying of the system.   
\end{abstract}

\vspace{5mm}


\section{Introduction}
\label{sec_intro}

The cytoplasm of living cells contains a complex network of fibres (the cytoskeleton) that helps to maintain cell shape by providing structural support, and that facilitates intracellular transport. This transport is mediated by specialized mechano-proteins, the so-called molecular motors, that utilize ATP to move vesicles, organelles (or other cargo) in a particular direction along the cytoskeleton \cite{Val_2003}. Microtubules (MTs) are one type of cytoskeletal element that consist of thirteen oriented protofilaments \cite{Tilney73} (that we refer as {\em lanes}), each of which can support bidirectional transport. Long-distance transport along MTs is powered by kinesin and dynein, where kinesin takes its cargo to the polymerization-active plus-end of a MT while dynein walks towards the minus-end \cite{Val_2003}.

Various lattice models have been proposed to understand the role of cooperation and competition between motors involved in such bi-directional transport. These are typically generalizations of asymmetric simple exclusion processes (ASEP) in one dimension \cite{Derrida_etal_1993,Kolomeisky1998} where  motors are represented by particles that move on a lattice. Already for unidirectional motor transport models \cite{Muller2005,Parmeggiani_etal_2003,Popkov_etal_2003} there are nontrivial effects such as traffic jams due to the mutual exclusion of particles on sites of the MT \cite{Muller2005,Lipowsky2001}.

Bidirectional transport has previously been modelled by an exclusion process with binding and unbinding events \cite{Klumpp2004,Muhuri2008,EbbSan09,EbbAppSan10} or a lattice model with site sharing \cite{Liu10}.  Change in direction of individual cargos may be due to ``tug-of-war'' \footnote{``tug-of-war'' events refer to simultaneous and competitive activity of counteracting motors on the same organelle.} events \cite{Mue_etal_2008,Mue_etal_2010,Hendricks_etal_2010,Soppina_etal_2009} and this can be incorporated into ASEP models by assuming that motion of a single type of particle in different directions takes place on different lanes (and change of direction is modelled simply by change of lane) \cite{Juh07,Juh10,Ashwin_etal_2010} or that particles may change type \cite{Schuster_etal_2010}. In the latter case, counter-moving particles approaching on the same lane can only pass if at least one of them can change lane. The lane-change rules in the latter model are based on experimental observations that (a) dynein can switch between lanes on the MT at a certain rate \cite{Wang_etal_1995}, whereas (b) kinesin remains on a single lane \cite{Ray1993}, and (c) no long-lasting traffic jam is observed between particles away from the tip. However, recent reports show that kinesin can change lanes to overcome obstacles on the MT \cite{Dreblow10}. Thus the situation in the living cell is less clear and one aim of this paper is to investigate the impact of different lane-change rules on bidirectional transport. 

We study a generalization of the models from \cite{Ashwin_etal_2010,Schuster_etal_2010} for bidirectional motion motivated by {\em in vivo} experimental observations of bidirectional motion of dynein particles near the hyphal tip of {\em Ustilago maydis}. The particles represent dynein that is transported towards the plus-end of a MT by kinesin-1 and towards the minus-end under its own power. As the dynamics of the MT in  this system are comparatively stable \cite{Steinber2001} and binding/unbinding events are rare in this system \cite{Schuster_etal_2011}, we focus on modelling bidirectional transport of particles that remain bound at all times to a fixed section of MT where one end is a plus-end (with no-flux boundary conditions are applied) while at the other end we inject particles at a given {\em injection rate}. In particular we explore the behaviour of the tip accumulation (or {\em tip size}) as a function of this injection rate.

The paper is organized as follows: In Section~\ref{sec_model} we introduce a general multi-lane model for bidirectional motion and demonstrate that it includes, as special cases, the models of \cite{Ashwin_etal_2010,Schuster_etal_2010}.
Section~\ref{sec_influence} examines the influence of collisions and lane changes on bidirectional transport in this model. In particular we find cross-lane diffusion due to lane changes, and for low injection rates we find an approximately linear scaling of the tip size with injection rate as discussed in \cite{Ashwin_etal_2010}. However, for higher injection rates we find there is a nonlinear growth in the tip size (depending on the lane-change rules) due to trapping of particles near the plus-end, and a singularity at a finite critical injection rate. We suggest that is associated with a phase transition of the system. We also study how the lane change rules influence the tip structure and discuss measures of efficiency for the bidirectional transport. Interestingly, we find that the {\em in vivo} parameters obtained from living cell images \cite{Schuster_etal_2010} are close to optimal for effective bidirectional transport, in that they balance efficient transport to the tip against the average delay at the tip. 
In Section~\ref{sec_critical} we describe novel {\em pulsing states} (with whole-system approximately periodic filling and emptying) and {\em filled states} that can both appear at injection rates beyond the critical value. We finish with a discussion of biological relevance, implications and generalizations of these results in Section~\ref{sec_shortdisc}.

\section{A multi-lane model for bidirectional transport}
\label{sec_model}

\begin{figure}[ht]
\centerline{\epsfig{file=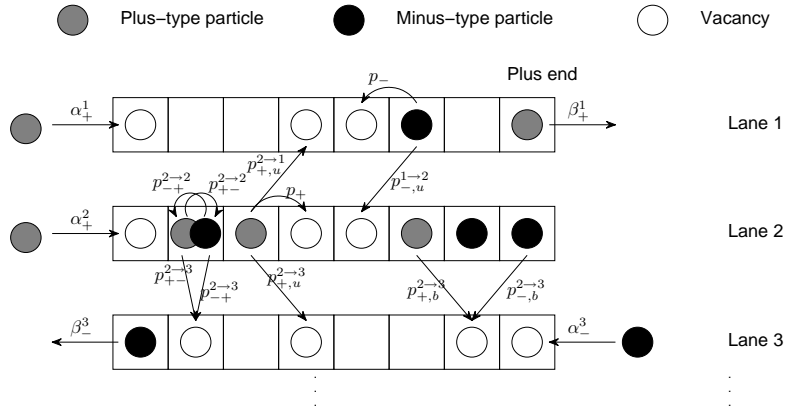,width=11cm}}
\caption{
Schematic diagram showing transition rates for the multi-lane bidirectional ASEP model; we consider $M$ lanes on the surface of a cylinder. The plus end (resp. minus end) of the MT is at the right (resp. left). Plus-type (resp. minus-type) particles move step forward with rate $p_+$ (resp. $p_-$) while they move forward associated with a change of lane with rates $p^{l\rightarrow l\pm 1}_{+,b(u)}$ (resp. $p^{l\rightarrow l\pm 1}_{\pm,b(u)}$) when blocked (resp. unblocked). Plus-type (resp. minus-type) particles are injected into the left (resp. right) boundary of the system at rate $\alpha^{l}_+$ (resp. $\alpha^{l}_-$) and exit at rate $\beta^{l}_+$ (resp. $\beta^{l}_-$) in the $l-$th lane. Particles can also change from plus-type to minus-type (resp. from minus-type to plus-type) with rate $p^{l\rightarrow k}_{+-}$ (resp. $p^{l\rightarrow k}_{-+}$) associated with a possible change of lane $l$ to $k$ (if $l\neq k$); we assume the site is preserved during a change of type.
}\label{fig_multi_asep}
\end{figure}

Consider a bidirectional transport model where particles move on a lattice of spatial locations consisting of $M$ adjacent lanes around the circumference of a cylinder.\footnote{We identify lane $M+1$ with lane $1$ to represent this cylinder.} Each oriented lane is discretized into $N$ {\em sites} between a {\em plus-end} and {\em minus-end}, as illustrated in Figure~\ref{fig_multi_asep}. Particles are of two types; {\em plus-type} particles move towards the plus-end, while {\em minus-type} particles move towards the minus-end; a single particle represents a bound pair of opposite-directed motor proteins that is pulled along the lane by one of the motors being bound to the MT and the type of the particle corresponds to which of the motors is currently bound to the MT. 

We identify each location along the cylinder by a pair $(l,i)$ where $l\in\{1,\cdots,M\}$ denotes the lane and $i\in\{1,\cdots,N\}$ the site along the lane and let $\tau^{l}_{\pm,i}=1$ or $0$ represent the presence or absence of a plus-type (minus-type) particle at location $(l,i)$. Each location is occupied by at most one particle (i.e. $\tau^{1}_{+,i}+\tau^{l}_{-,i}$ can only be $0$ or $1$) and particles move from one location to another at given rates. We also allow the possibility that each particle can change from one type to the other at rates representing the resolution of brief ``tug-of-war'' events \cite{Mue_etal_2008,Mue_etal_2010,Hendricks_etal_2010,Soppina_etal_2009} between opposite oriented motor proteins bound to the particle. This change of type may or may not be associated with a change of lane. 

The plus- and minus- type  particles may have transition rates that describe {\em motion} and {\em particle-type change} and both of these may depend on site; however, here we will assume that they are independent of site, though they may depend on lane. We say a model is {\em lane-inhomogeneous} (or simply {\em inhomogeneous}) if the transition rates and/or boundary conditions are not uniform between lanes; otherwise we say the model is (lane) {\em homogeneous}. A particle is {\em blocked} if there is another particle that prevents it from undergoing the {\em forward motion} below, otherwise it is {\em unblocked}. Transition rates may depend on whether a particle is blocked or not as illustrated in Figure~\ref{fig_multi_asep} and the possible transitions we consider are defined below:
\begin{itemize}
\item {\bf Motion:}
Plus-type particles move from $(l,i)$ to $(k,i+1)$ with a change of lane $l$ to $k$ when unblocked (resp. blocked) at rate $p^{l\rightarrow k}_{+,u}$ (resp. $p^{l\rightarrow k}_{+,b}$). Similarly, minus-type particles move from $(l,i)$ to $(k,i-1)$ at rate $p^{l\rightarrow k}_{-,u(b)}$. Motion is subject to an exclusion principle -  a change can only occur if the target location to move into is vacant. {\bf Forward motion} is used to mean the motion on a single lane (i.e., $k=l$) in either direction; this can occur only to a particle that is unblocked. If this forward motion is lane-homogeneous then we write
$$
p_+:=p^{l\rightarrow l}_{+,u},~~~~p_-:=p^{l\rightarrow l}_{-,u}.
$$
\item {\bf Particle type/direction change:}
Plus-type particles can change to minus-type particles with a change of lane $l$ to lane $k$ at rate $p^{l\rightarrow k}_{+-}$. Minus-type particles can change to plus-type particles with a similar change in lane at rate $p^{l\rightarrow k}_{-+}$ (we assume that the site $i$ is preserved for a change in type). For lane-homogeneous direction changes on the same lane, we write
$$
p_{+-}:=p^{l\rightarrow l}_{+-},~~~p_{-+}:=p^{l\rightarrow l}_{-+}.
$$
\end{itemize}
Boundary conditions for the model are assumed as follows:
\begin{itemize}
\item {\bf Inflow:}
Plus-type (resp. minus-type) particles are injected at rate $\alpha^l_+$ (resp. $\alpha^l_-$) into the minus end (reps. plus end) of the $l-$th lane. Both cases are subject to an exclusion principle.
\item {\bf Outflow:}
Plus-type (resp. minus-type) particles exit from the plus end (resp. minus end) of the $l-$th lane at rate $\beta^l_+$ (resp. $\beta^l_-$).
\end{itemize}

The state of the model $\{\tau^{l}_{\pm,i}(t)\}$ at time $t$ changes according to the above events (assumed to take place independently and instantaneously) from an initial state $\{\tau^{l}_{\pm,i}(0)\}$. In summary, the model for a given lattice (fixed $M$ and $N$) has a number of parameters
$$
p^{l\rightarrow k}_{\pm,b},~~p^{l\rightarrow k}_{\pm,u},~~p^{l\rightarrow k}_{+-},~~p^{l\rightarrow k}_{-+}, ~~\alpha^l_{\pm}~~\mbox{ and }~~\beta^l_{\pm}
$$
for $k,l=1,\cdots M$ that collectively determine the bidirectional transport behaviour. We write $\alpha_+=\sum_{l=1}^{l=M} \alpha_+^l$ to denote the total inflow of plus-type particles. In practice, many of these rates will be zero - for example, here we only permit a type change of a particle that remains on the same lane or moves to an adjacent lane. As in other ASEP models, we assume that the dynamics for typical choices of parameters converges to a unique statistical equilibrium independent of initial conditions, though possibly after a long transient. Exceptions occur if there is a degeneracy of the parameters, for example if no lane-changes are permitted, i.e., $p^{l\rightarrow k}_{\pm,b(u)}=0$ for $k,l=1,\cdots M$, then accumulations can be created at arbitrary locations that are never cleared. The system exhibits various different dynamical regimes depending on parameters. We have not attempted to characterize the phase diagram of these states in full, but do discuss this further in Section~\ref{sec_pulsing}.

The equilibrium densities of plus- and minus-type particles are defined to be the mean occupancy of the particles:
\begin{equation}
\rho_i^l:=\langle \tau_{+,i}^l(t)\rangle,~~\sigma_i^l:=\langle\tau_{-,i}^l(t)\rangle
\end{equation}
where $\langle\cdot\rangle$ denotes the ensemble average; assuming ergodicity this can be evaluated using a time average. In particular, if there is an accumulation at the tip then we denote the {\em lane tip length} $\lambda^{l}_{tip}(t)$ as the number of particles in the accumulation within lane $l$ at time $t$, and the {\em tip size} $n_{tip}(t)$ is
$$
n_{tip}(t)= \sum_{l=1}^M \lambda^{l}_{tip}(t).
$$
with mean tip size as $\left<n_{tip}\right>$ at a steady state.

There are few methods giving exact analytical solutions of the equilibrium state for ASEP models (we refer to \cite{BlytheEvans_2007} for a review of ASEP models and their analytical solutions). Therefore, we numerically simulate this discrete event, continuous time model using both Gillespie \cite{Gillespie1977} and fixed time-step ($h_t$) Monte Carlo methods. The latter method converges to the Gillespie method and gives outcomes that are independent of update methods in the limit $h_t\rightarrow 0$. This general multi-lane model can reduce to previously studied ASEP models in special cases. For instance, restricting to $M=1$ and one particle-type  gives the simplest unidirectional ASEP model. In particular of bidirectional transport, this multi-lane model limits the two models recalled in the following sections. 

\subsection{A simple two-lane model}\label{sec_twolane}

\begin{table}
\caption{\label{tab_twolane}
The transition rates (a) with boundary conditions I in (b) for a two-lane model ($M=2$) that gives bidirectional transport where each lane supports transport in one direction only and allows lane changes only together with a type change; if also associated with boundary Conditions II, the model corresponds to no-flux boundary at the plus end.
}
\begin{indented}
\item[]\begin{tabular}{@{}l}
\br
(a)  Transport rates \cr
\mr
~~ $p^{l\rightarrow l}_{-+}=p^{l\rightarrow l}_{+-}=0,~~p^{l\rightarrow l\pm 1}_{\pm, u(b)}=0 ~(l=1,2),~~
p_d:=p^{1\rightarrow 2}_{+-},~~p_u:=p^{2\rightarrow 1}_{-+},~~p_{\pm}:=p_{\pm,u}^{l\rightarrow l}$\cr
\mr
\end{tabular}
\\
\begin{tabular}{@{}llll}
(b) & Boundary conditions I & & Boundary conditions II\cr
\mr
& $\alpha^2_{+}=\alpha^1_{-}=0$ & &$\beta^1_{+}=\alpha^2_{-}=0$\cr
\br
\end{tabular}
\end{indented}
\end{table}

A lattice model for bidirectional transport needs to either segregate, or deal with collisions between, opposite-directed particles. Works such as \cite{Klumpp2004,EbbAppSan10} include the possibility of binding/unbinding from/to a reservoir in their model. In our multi-lane model, for $M=2$ one can segregate the particles into two lanes of unidirectional motion by setting the parameters in Table~\ref{tab_twolane}~(a) with boundary conditions I as in Table~\ref{tab_twolane}~(b). This particular choice of parameters corresponds to the two-lane model of \cite{Juh07,Juh10} and segregates the particles into two lanes- one lane will have no particles of minus-type, while the other lane will have no particles of plus-type, i.e., the equilibrium densities $\rho^2=\sigma^1=0$. If, in addition, boundary conditions II in Table~\ref{tab_twolane}~(b) are satisfied then there is no flux at the right boundary and the model corresponds to the special case studied in \cite{Ashwin_etal_2010}. We recall from \cite{Ashwin_etal_2010} that in these circumstances, if the minus-end boundary condition $\alpha^1_+=\alpha_+$ small enough on the first lane and $\beta^2_-=p_-$ on the second lane then a shock forms near the tip that traps a number of particles there. Using a mean-field approximation and denoting $x=i\delta$ where $\delta=1/N$, one can find an asymptotic profile for the equilibrium densities of $\rho$ and $\sigma$ for plus- and minus-type particles along the domain as follows (assuming $\sigma^2$ and $\min\{\rho^1, (1-\rho^1)\}$ are of order $\delta$):
\begin{eqnarray}
\rho^1(x) &\approx & \left\{\begin{array}{ll}
\frac{\alpha_+}{p_+}
\exp\left[x\left(\frac{Np_{u}}{p_-}-\frac{Np_{d}}{p_+}\right)\right], &x\in[0,x_s]\\
 1+\frac{p_{d} N}{p_+}(x-1), &x\in [x_s,1]  
\end{array}\right.,\label{eq_den_twolane}\\
\sigma^{2}(x) &= &\left\{\begin{array}{ll}
\rho^1(x), &x\in[0,x_s]\\
 1-\rho^1(x), &x\in [x_s,1]  \end{array}\right.,\label{eq_den_twolane_minus}
\end{eqnarray}
where $x_s$ represents the shock position. Continuity of the flux \cite{Ashwin_etal_2010} then implies that the {\em mean tip size} can be approximated as
\begin{equation}\label{eq_ntip}
\left<n_{tip}\right>\approx (1-x_s)N\approx\frac{\alpha_+}{p_d}\exp\left(\frac{Np_{u}}{p_-}-\frac{Np_{d}}{p_+}\right).
\end{equation}

\subsection{A homogeneous thirteen-lane model}
\label{sec_thirteenlane}

Taking $M=13$ and considering lane-homogeneous transition rates as in Table~\ref{tab_thirteen}~(a,b), the multi-lane model reduces to the thirteen-lane model of \cite{Schuster_etal_2010}. If we consider boundary conditions
\begin{equation}\label{eq_boundary}
\alpha^l_{+}=\alpha_+/M,~~ \alpha^l_{-}=0,~~\beta^l_+=0~~\mbox{ and }~~\beta^l_-=p_+
\end{equation}
for all $l\in\{1,\cdots,M\}$, this corresponds to homogeneous injection of plus-type particles into the minus-end and no flux at the plus-end of the domain, while minus-type particles exit without impediment. Experiments and other information reported in \cite{Schuster_etal_2010} suggest the rates listed in Table~\ref{tab_thirteen}~(a) are appropriate for this system. Rates in Table~\ref{tab_thirteen} and boundary conditions in (\ref{eq_boundary}) are considered as default parameters for simulations unless otherwise specified. 

\begin{table}[ht]
\caption{\label{tab_thirteen}
Transition rates for the thirteen-lane model based on {\em in vivo} experiment measurements or estimations of velocities, mean run length and fluxes on a MT for the {\em Ustilago maydis} hyphal tip as detailed in \cite{Schuster_etal_2010}. Note that $\alpha_+$ represents the total injection rate, i.e., $\alpha_+=\sum_{l=1}^{M}\alpha_+^l$. Simulations use $N=1250$ sites (i.e., $L=10$ $\mu m$ in length discretised with spatial step $h_s=8$ $nm$) and a sequential Monte-Carlo update with time step $h_t=0.0042$ s unless otherwise specified.
}
\begin{indented}
\item[]\begin{tabular}{@{}l}
\br
(a) Possible transition rates $s^{-1}$ \cr
\begin{tabular}{@{}llllllllll}
\hline
$p_{+}$ & $p_-$ & $p_{+-}$& $p_{-+}$ & $p^{l\rightarrow l\pm 1}_{+,u}$ & $p^{l\rightarrow l\pm 1}_{+,b}$ & $p^{l\rightarrow l\pm 1}_{-,u}$ & $p^{l\rightarrow l\pm 1}_{-,b}$ & $\alpha_+$ & $\beta_-^l$\cr
212.5  & 203.83  & 0.0406 & 0.0273  & 0 & 0  & 4.335  & 106.25 & 1.06 & 212.5\cr
%\hline
\end{tabular}
\\
\begin{tabular}{@{}llll}
\br
(b)&  Only adjacent lane changes permitted && Type changes occur on a same lane\cr
\hline
& $p^{l\rightarrow k}_{\pm,b(u)}=0 ~(k\neq l,l\pm 1)$&&  $p^{l\rightarrow k}_{-+}=p^{l\rightarrow k}_{+-}=0 ~(k\neq l)$\cr
\br
%\hline
%\end{array}
%$$
\end{tabular}
\end{tabular}
\end{indented}
\end{table}

\section{Influence of collisions and lane-changes on transport properties}\label{sec_influence}

In this section, we consider the multi-lane model with lane-homogeneous transition rates satisfying Table~\ref{tab_thirteen}~(b) and boundary conditions as in (\ref{eq_boundary}) except for possibly inhomogeneous injection rates. For lane-homogeneous and symmetric lane-change rates, we remove the superscript and write
$$
p_{+,u(b)}:=p^{l\rightarrow l\pm 1}_{+,u(b)};~~~p_{-,u(b)}:=p^{l\rightarrow l\pm 1}_{-,u(b)}
$$
independent of lane $l$.

\subsection{Mean field approximation and cross-lane diffusion}
\label{sec_crosslane}

In cases of inhomogeneous boundary conditions, the densities of plus- or minus-type particles can be homogenised by lane changes of particles, and lane changes are necessary for this to happen. Depending on which particles change lane under which circumstances (blocked or unblocked), this homogenization can occur to one or both types of particle. 

\begin{figure}
\centerline{\epsfig{file=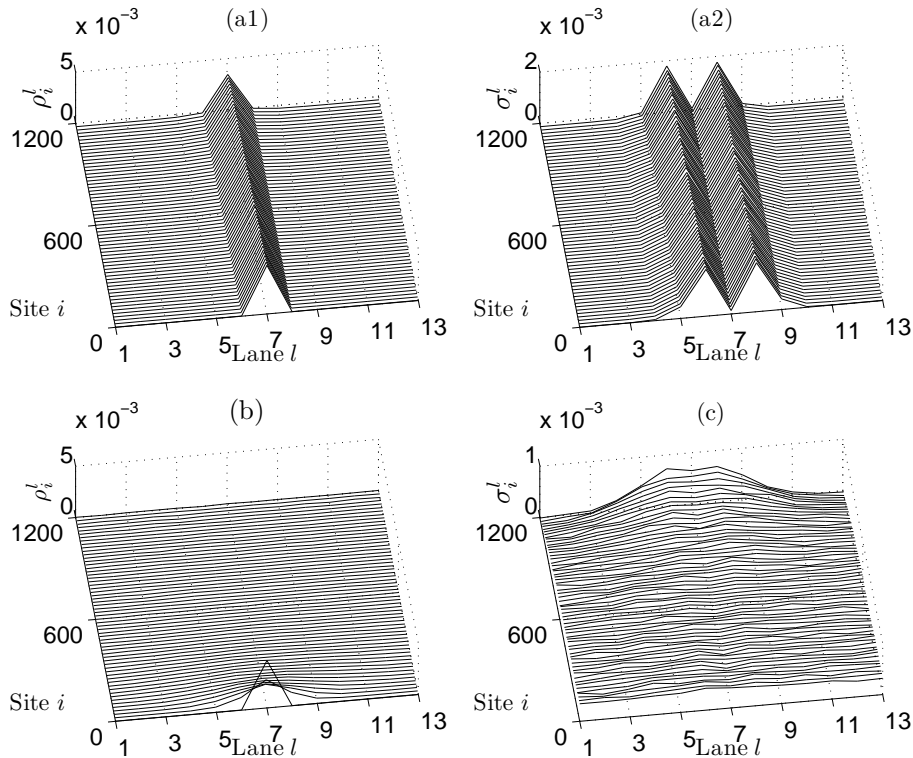,width=14cm}}
\caption{
Densities of plus- or minus-type particles corresponding to default parameters in Table~\ref{tab_thirteen} and boundary conditions in~(\ref{eq_boundary}) except that the lane-change rates when unblocked are varied and there is injection  only in one lane (namely $\alpha^{7}_+=1~s^{-1}$ and $\alpha^{l}_+=0$ for $l\neq 7$). (a1) and (a2) show the inhomogeneous densities of plus- and minus-type particles respectively, when $p_{\pm,u}=0$, i.e., particles only change lanes after collisions; (b) (resp. (c)) shows cross-lane diffusion leading to a more homogeneous density of plus-type (resp. minus-type) particles away from the boundaries with lane-change rates when unblocked as $p_{-,u}=0,~p_{+,u}=4.335~s^{-1}$ (resp. $p_{-,u}=0.4335~s^{-1},~p_{+,u}=0$). In all cases the densities near the tip (not shown) are high.
}
\label{fig_dendiffu}
\end{figure}

For example, if only minus-type particles can change lanes when unblocked in a dilute situation, then the density of minus-type becomes lane-homogeneous by cross-lane diffusion (see Figure~\ref{fig_dendiffu}~(c)), while the density of plus-type particles (not shown) may still be inhomogeneous due to the inhomogeneous injection rates and no lane changes of this type. Even if only one type of particles changes lanes when unblocked, the density of both types of particles will be smoothed as shown in Figure~\ref{fig_dendiffu}~(b) for the density of particle-type particles. Exceptions to this are shown in Figure~\ref{fig_dendiffu}~(a1,a2) when lane change only occurs after collision - in this case the plus-type particles injected into the middle lane ``sweep'' that lane clear of minus-type particles.

These effects can be understood in terms of cross-lane diffusion of particles induced by lane changes in a dilute region. If we apply the mean field description of~\ref{app_meanfield} to a dilute region and assume small variation of density with site $x=i \delta$ ($x\in[0,1]$ and $\delta=1/N$), in the case of $p_{-,u}=p_{+,u}=0$, the solutions~(\ref{equ_rhoequl0}) and (\ref{equ_sigmaequl0}) in~\ref{sec_inhom} indicate the non-homogeneity of densities in both types of particles if given non-homogeneous injection rates $\alpha^l_+$. If we allow lane changes when unblocked and assume that a change of type occurs more slowly than a change of lane, i.e., $p_{-+},p_{+-}\ll p_{\pm,u}$, then the stationary state distribution $\rho^l(x)$ satisfies (see~\ref{sec_inhom} for details):
$$
0=-\delta p_+\frac{d\rho^l}{dx} + p_{+,u} \left[ \rho^{l-1}+\rho^{l+1}-2\rho^l-\delta \frac{d\rho^{l-1}}{dx}-\delta\frac{d\rho^{l+1}}{dx}\right],~~l=1,2,\cdots,M
$$
to leading order in $\delta$. Solving this differential-difference equation for $M$ lanes gives
$$
\rho^l(x) = \sum_{k=0}^{\lfloor M/2 \rfloor} \left[A_k \cos \frac{2\pi kl}{M}+B_k \sin \frac{2\pi kl}{M}\right]\exp \left[ - \kappa_+(k) x\right]
$$
where
$$
\kappa_+(k)= \frac{2p_{+,u}\left[1-\cos\frac{2\pi k}{M}\right]}{\delta p_+ + 2p_{+,u}\delta\cos\frac{2\pi k}{M}},
$$
and $A_k,B_k$ are constants determined by the density at $\rho^l(x_0)$. Similarly, the density for the minus-type particles can be approximated as
$$
\sigma^l(x) = \sum_{k=0}^{\lfloor M/2 \rfloor} \left[A'_k \cos \frac{2\pi kl}{M}+B'_k \sin \frac{2\pi kl}{M}\right]\exp \left[\kappa_-(k) x\right]
$$
where
$$
\kappa_-(k)= \frac{2p_{-,u}\left[1-\cos\frac{2\pi k}{M}\right]}{\delta p_- + 2p_{-,u} \delta\cos\frac{2\pi k}{M}},
$$
and $A'_k,B'_k$ are constants determined by $\sigma^l(x_0)$.
This suggests, not surprisingly, that the main effect of the lane-change in dilute cases is simply a diffusion of the density across the lanes in the direction of travel.

Assuming there is a unique equilibrium state, the density for plus- and minus-type particles in the multi-lane model with homogeneous rates will have lane homogeneity though it will depend on site. For small injection rate $\alpha_+$, the system tends to a stationary state that is dilute near the minus end and in a high density near the plus end. Using a similar method as detailed in~\ref{app_meanfield} to the lane-homogeneous case \cite{Ashwin_etal_2010}, for small injection rate, under the conditions that $\rho,\sigma=O(\delta)$ far away from the tip and ignoring $O(\delta^2)$, we find similar approximate expressions as~(\ref{eq_den_twolane}) for the stationary state in dilute regions; see (\ref{equ_den_13lane}). For large enough $\alpha_+$, dynamically {\em pulsing states} or {\em filled states} as discussed in Section~\ref{sec_pulsing} will appear.

\subsection{A phase transition in the tip accumulation}
\label{sec_tipsize}

For the half-closed system (i.e. closed at the plus end) in both the two-lane \cite{Ashwin_etal_2010} and thirteen-lane models \cite{Schuster_etal_2010}, it is clear that for typical particle-type change rates (e.g., $p_{+-}\ll p_+$), particles accumulate at the tip. On the other hand, from the mean tip size approximation~(\ref{eq_ntip}) for the two-lane model, we see that $\left<n_{tip}\right>$ increase linearly with injection rate $\alpha_+$. However, this is not necessarily the case for the multi-lane (say, $M=13$) model. In the following, we investigate the size of the tip accumulation and its dependence on the total injection rate $\alpha_+$ for the thirteen-lane model with otherwise default parameters as in Table~\ref{tab_thirteen} and boundary conditions as in (\ref{eq_boundary}).

\begin{figure}[ht]
\begin{minipage}{0.5\textwidth}
\centering
\includegraphics[scale=0.45]{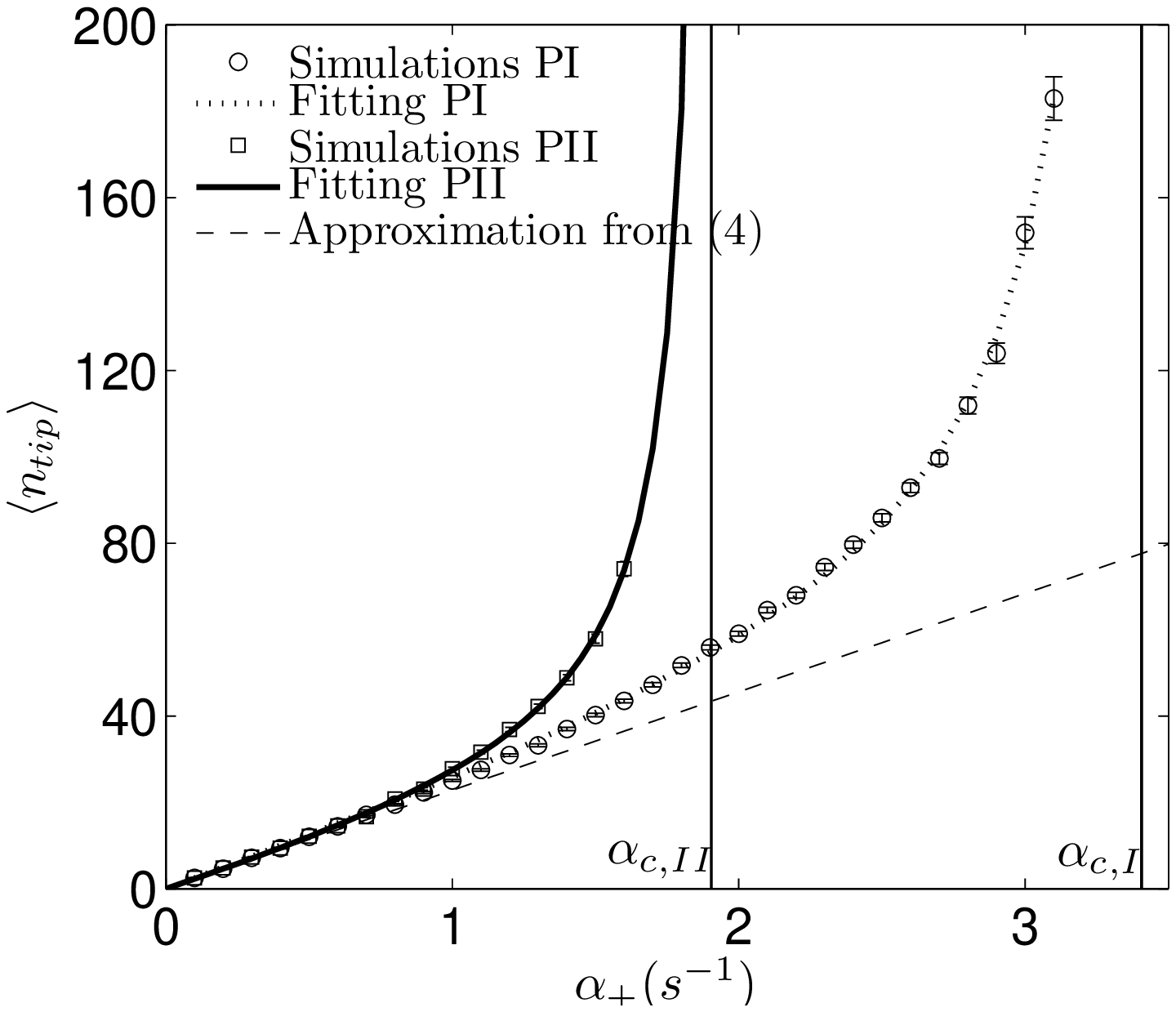}\\
(a)
\end{minipage}
\begin{minipage}{0.5\textwidth}
\centering
\includegraphics[scale=0.45]{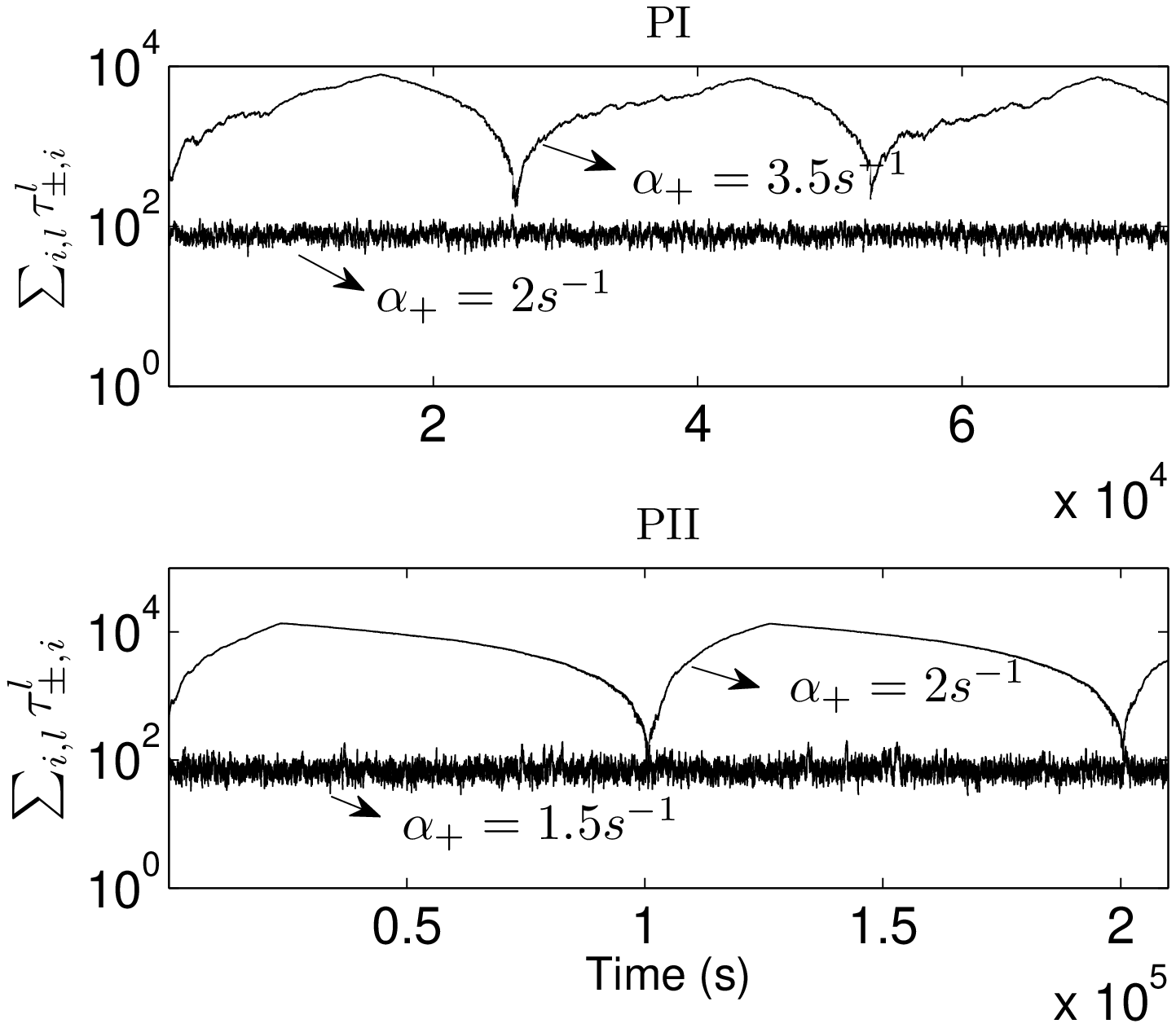}\\
(b)
\end{minipage}
\caption{
(a) shows mean tip size as a function of total injection rate $\alpha_+$ for the multi-lane (M=13) model. Simulations PI refers to lane-change protocol PI where $p_{-,b}=p_+/2$ and $p_{+,b}=0$ while simulations PII is for protocol PII where $p_{\pm,b}=p_+/2$.
In both protocols we find a good fit to a rational function of $\alpha_+$ with a singularity at a critical value $\alpha_{c,I}\approx 3.406~s^{-1}$ (fitting PI) or $\alpha_{c,II} \approx 1.947~s^{-1}$ (fitting PII). For comparison, the straight dash line shows the linear dependence~(\ref{eq_ntip}) of $\alpha_+$ in the two-lane model recalled in Section~\ref{sec_twolane}; other parameters use the values of corresponding rates in Table~\ref{tab_thirteen}; see \cite{Ashwin_etal_2010} for details. (b) shows the total occupancy $\sum_{l,i}\tau_{\pm,i}^l$ as a function of time for two typical injection rates as indicated, beyond and below the critical value for each of the lane-change protocols PI and PII. Other parameters used in the multi-lane model are default as in Table~\ref{tab_thirteen} and boundary conditions as in (\ref{eq_boundary}).
}
\label{fig_tipsize}
\end{figure}

We compare two lane-change protocols PI and PII (further protocols are considered in the next section). The protocol PI only allows minus-type particles to change lanes when blocked while protocol PII allows both type to change lanes when blocked; lane-change rates are assumed homogeneous and symmetric (see details in Section~\ref{sec_tipstru}). Simulations gave a mean tip size that depends on $\alpha_+$, fitting to a rational function with three fitting parameters $A,B$ and $C$
$$
\left<n_{tip}(\alpha_+)\right>=\frac{\alpha_+A(1+\alpha_+C)}{1-\alpha_+B};
$$
see Figure~\ref{fig_tipsize}~(a) and Table~\ref{tab_nlinfit}. Note that there is a singularity in this rational fitting function at $\alpha_+=1/B$.
For the simulated system, these critical values are $\alpha_{c,I}=3.406~s^{-1}$ and $\alpha_{c,II}=1.947~s^{-1}$ respectively for lane-change protocol PI and protocol PII. Moreover, by comparing the change of the total occupancy of particles in the system in time between injection rates below and beyond the critical value, Figure~\ref{fig_tipsize}~(b) indicates the existence of a transition between a {\em shock state} (where there is an approximately steady accumulation at the tip) and a new phase of motion, an unsteady {\em pulsing state} which is further examined in Section~\ref{sec_pulsing}. 

Additionally, by comparison with approximation~(\ref{eq_ntip}) of the mean tip size from the two-lane model, we note that the mean tip size in the multi-lane model increases initially linearly but then nonlinearly for both lane-change protocols. This nonlinear increase is due to trapped minus-type particles at the tip and is investigated in the following section.

\begin{table}
\caption{
Best fit of the data from simulations in Figure~\ref{fig_tipsize}~(a) to a rational function $\left<n_{tip}(\alpha_+)\right>=\frac{\alpha_+A(1+\alpha_+C)}{1-\alpha_+B}$ with parameters $A,B,C$. The values indicate best fit values and standard errors. 
}
\begin{indented}
\item[]\begin{tabular}{@{}c|ccc}
\br
& A & B & C \cr
\hline
PI &  24.45$\pm$ 0.329 & 0.2936$\pm$ 0.00139 &-0.2525$\pm$ 0.00388 \cr
PII & 22.18$\pm$ 0.509 & 0.5136 $\pm$ 0.00908 &-0.3978$\pm$ 0.002219 \cr
\br
\end{tabular}
\end{indented}
\label{tab_nlinfit}
\end{table}

\subsection{Influence of lane-changes on the tip accumulation}
\label{sec_tipstru}

The size of accumulation at the tip clearly depends on the lane-change rules; see Figure~\ref{fig_tipsize}~(a). Recall from Section~\ref{sec_crosslane} that lane changes can homogenize the densities so we expect the homogeneity of $\lambda^l_{tip}(t)$ among the lanes will depend on the lane-change rules when blocked. To confirm this, we examine the maximum difference among the $\lambda^l_{tip}(t)$ defined as
$$
\Delta \lambda_{tip}(t)=\max\{ \lambda^{l}_{tip}(t)-\lambda^{k}_{tip}(t) :k,l=1,\cdots, M\}.
$$
The distribution of this quantity gives a measure that characterizes how ``smooth'' ($\Delta\lambda_{tip}$ small) or ``ragged''  ($\Delta\lambda_{tip}$ large) the tip is. In particular, the larger the mean value $\left<\Delta \lambda_{tip}\right>$, the easier it is for minus-type particles to be released from the tip.

\begin{figure}[ht]
\centerline{\epsfig{file=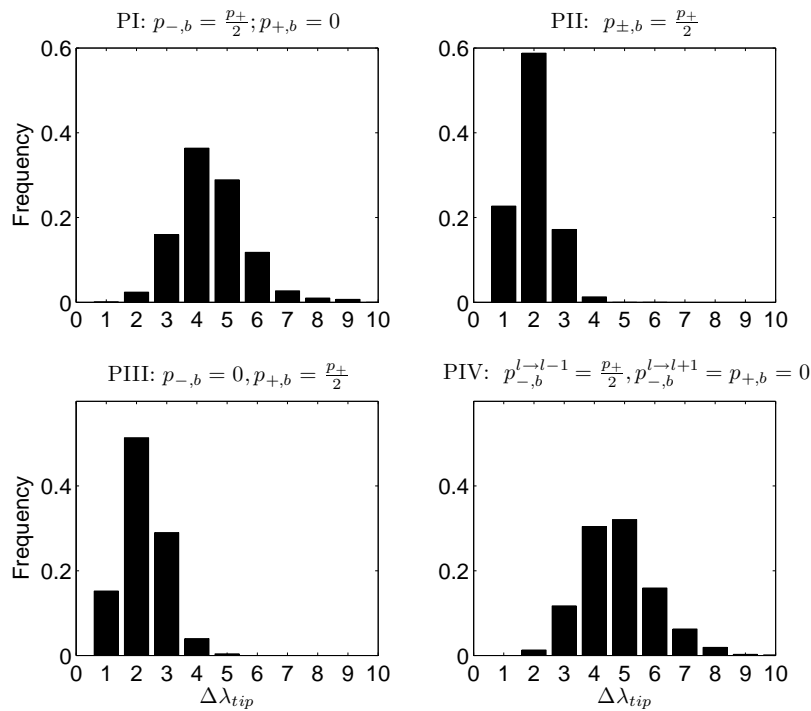,width=12cm}}
\caption{
Distribution of maximum length difference $\Delta\lambda_{tip}$ for lane-change protocols PI-IV; other parameters are default as in Table~\ref{tab_thirteen} and boundary conditions as in~(\ref{eq_boundary}). Note that the structure of the tip accumulation in PI or PII is clearly smoother than that in PI and PIV.
}
\label{fig_dif}
\end{figure}

We consider four different lane-change (when blocked) protocols with homogeneous lane-change rates in an attempt to better understand their influence on the structure of tip accumulation via the distribution of $\Delta\lambda(t)$ at stationary state. These are:
\begin{itemize}
\item[(PI)] $p_{-,b}=p_+/2$, $p_{+,b}=0$ -- only minus-type particles are allowed to change lanes with a rate (homogeneous and symmetric) that preserves the velocity; this corresponds to the assumption used in \cite{Schuster_etal_2010};
\item[(PII)] $p_{\pm,b}=p_+/2$ -- both minus- and plus-type particles are allowed to change lanes;
\item[(PIII)] $p_{-,b}=0$, $p_{+,b}=p_+/2$ -- only plus-type particles are allowed to change lanes;
%\item[(PIV)] $p^{\uparrow}_{-,b}=p_+/2$, $p^{\downarrow}_{-,b}=0=p_{+,b}$ -- only minus-type particles are allowed to change and to only one of the adjacent lanes.
\item[(PIV)] $p^{l\rightarrow l-1}_{-,b}=p_+/2$, $p^{l\rightarrow l+1}_{-,b}=0=p_{+,b}$ -- only minus-type particles are allowed to change and to only one of the adjacent lanes with lane-homogeneous rate.
\end{itemize}
We ignore lane changes in the unblocked case as these are clearly not significant for escape from the tip.  The distributions of the length differences $\Delta \lambda_{tip}$ at a fixed stationary time are shown in Figure~\ref{fig_dif} for above four different lane-change protocols; other rates are as default ones. Note that allowing plus-type particles to undertake lane changes ``smooths'' the structure of the tip. The results for lane-change protocols PI and PII agree with the homogenized density of both type of particles on the effect of lane changes (see Figure~\ref{fig_dendiffu}). Note the similar distribution of $\Delta_{tip}$ between PI and PIV (PII and PIII) from Figure~\ref{fig_dif}, it is then reasonable to consider protocols PI and PII as typical lane-change protocols.

For both lane-change protocols PI and PII, particles can be trapped under several layers in the tip accumulation. Shown in Figure~\ref{fig_tipsize}~(a) that the mean tip size increases more rapidly with injection rates for protocol PII, we suggest this is due to there being more minus-type particles ``trapped'' in the tip which in turn lead to larger increase on the tip size. We say a particle is ``trapped'' if its motion (including forward motion and lane changes) is obstructed by occupation of other particles and loss the ability to take the motion due to the exclusion principle. We define two fractions: the {\em trapped minus-type fraction} $F_{-,trap}$ and the {\em minus-type fraction} $F_{-}$ within the tip in the steady state as
$$
F_{-,trap}=\frac{\mbox{mean number of ``trapped'' minus-type particles at the tip}}{\mbox{mean number of total minus-type particles at the tip}},
$$
$$
F_{-}=\frac{\mbox{mean number of total minus-type particles at the tip}}{\left<n_{tip}\right>}.
$$
In the two-lane model discussed in Section~\ref{sec_twolane} where two types of particles move on different lanes, the density profile of minus-type particles is dilute, which indicates $F_{-,trap}\approx 0$. Additionally, from expressions~(\ref{eq_den_twolane}) and (\ref{eq_den_twolane_minus}), 
$$
F_-\approx\int_{x_s}^{1}\sigma(x)dx/\int_{x_s}^1(\rho(x)+\sigma(x))dx\approx \frac{p_d}{2p_+}\left<n_{tip}\right>
$$ 
For the particular parameters in Table~\ref{tab_thirteen} with $p_d=p_{+-}$, the minus-type fraction $F_-\approx 0.2\%$ associated with a mean tip size $\left<n_{tip}\right>\approx 20$. However, in the thirteen lane model, from Table~\ref{tab_trapped}, for low $\alpha_+=0.8~s^{-1}$ in lane-change protocol PI, $F_-\approx 4\%$ associated with $\left<n_{tip}\right>\approx 20$ and $F_{-,trap}\approx 40\%$. This shows a difference between the two-lane model and the thirteen-lane model even for low injection rates, though with a similar mean tip size. Moreover, in the two-lane model, viewed as a queueing process, the expected delay for particles at the tip is $1/p_{+-}$, while in the multi-lane model, the positive fraction of trapped minus-type indicates that the average time spent at the tip is larger then $1/p_{+-}$. This {\em average tip delay} will be discussed later on in Section~\ref{sec_efficiency}.

Comparing between the two lane-change protocols from Table~\ref{tab_trapped}, protocol PI gives lower $F_{-,trap}$ and $F_-$ than those in protocol PII, with the same $\alpha_+$ or the same mean tip size. This agrees with that a smoother structure of the tip accumulation is more likely to trap minus-type particles at the tip and indicates that the tip size and the (trapped) minus-type fraction are affected by each other. 

\begin{table}
\caption{\label{tab_trapped}
Comparison of trapped minus-type fraction $F_{-,trap}$ and minus-type fraction $F_-$ at the tip between lane-change protocols PI and PII. This shows for protocol PI that there are lower $F_-$ and $F_{-,trap}$ even for a similar mean tip size. When $\alpha_+ >\alpha_{c}$ in both protocols, $F_{-,trap}\approx 100\%$ and mean tip size does not convergent with time; marked as ``$-$''. Other parameters are default as in Table~\ref{tab_thirteen} and boundary conditions as in (\ref{eq_boundary}).
}
\begin{indented}
\item[]\begin{tabular}{@{}l|c|cccccccc}
\br
& $\alpha_+ (s^{-1})$& 0.8 & 1 & 1.2 & 1.4 & 1.6 & 1.8 & 2.0 & 4 \cr
\mr
& $F_{-,trap}$ & 37.8\% & 49.0\% & 66.7\% & 72.5\% & 79.6\% & 85.7\% & 88.2\% & 99.7\%\cr 
PI & $F_{-}$ & 3.7\% & 4.5\%  & 5.7\% & 7.6\% & 9.4\% & 12.1\% & 14.3\% & 51.3\% \cr
&  $\left<n_{tip}\right>$ & 19.7 & 24.9  &31.0 & 36.8 & 43.7 & 52.1 & 59.9 & --\cr
\mr
& $F_{-,trap}$ & 59.7\% & 83.3\% & 89.8\% & 92.6\% & 95.8\% & 98.5\% & 99.6\% & 100\% \cr
PII & $F_{-}$ & 5.4\% & 9.9\%  & 14.3\% & 20.2\% & 28.6\% & 42.7\% & 50.4\% & 57.6\%  \cr
 & $\left<n_{tip}\right>$ & 20.3 & 28.1  &  36.5 & 48.3 & 70.1 & -- & -- & --\cr
\br
\end{tabular}
\end{indented}
\end{table}

\subsection{Measures of bidirectional transport efficiency}
\label{sec_efficiency}

The previous subsections indicate that allowing only minus-type particles to change lanes as a result of collisions is more effective (with less trapped minus-type particles) than allowing both types of particles to change lanes. To understand more on the bidirectional transport observed in living cells, it is interesting to speculate on the implications of the {\em in vivo} experiment data, particularly in Table~\ref{tab_thirteen}, in terms of some non-trivial quantifiable efficiency.

There are many possible ways to quantify efficiency of bidirectional transport, depending on what sort of behaviour is important. For example, in some circumstances it could be that maximum speed is required while in other circumstances it could be that the maximum flux is required - for a practical system of bidirectional transport on a MT, different notions may be needed for different purposes. In a fungal model system, the particles accumulate at the MT plus end are thought to prevent endosome falling off the MT and to support their backward transport \cite{Schuster_etal_2010}. Therefore, efficiency transport of dynein requires a certain proportion of dynein reaching the tip to form an accumulation and a short duration at the tip. Therefore, we consider {\em particle arrival efficiency} and {\em average tip delay} defined  as following. 

\begin{figure}[ht]
\centerline{\epsfig{file=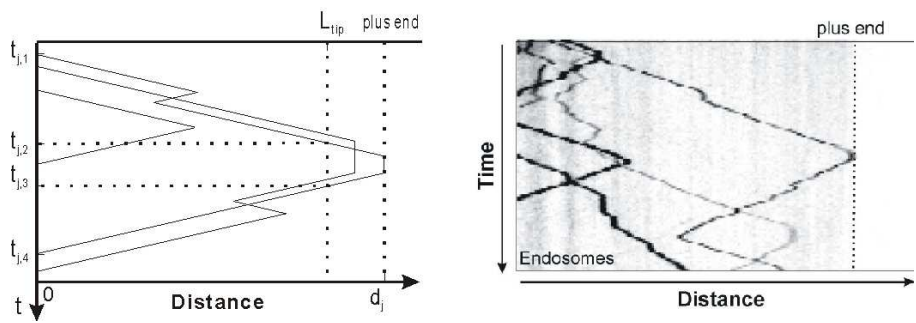,width=12cm}}
\caption{
Typical motion of three particles injected into the system (left panel). We show the times $t_{j,m}$ $m=1,\ldots,4$ for the $j-$th particle. Particles that exceed $L_{tip}$ are said to have entered the tip region. Kymograph (right panel) shows motility of early endosomes that were visualized by photo-activation of the endosome marker paGFP-Rab5a \cite{Schuster_etal_2010} in subapical regions of a cell of the fungus Ustilago maydis. Note the similarities in the two panels on the trajectories of motions.
}
\label{fig_eff_time}
\end{figure}

Suppose that the $n$-th particle injected at the minus end is at location $(l,i)=(L_n(t),I_n(t))$ at time $t$, and suppose that:
\begin{itemize}
\item it enters the system at time $t_{n,1}$ and leaves the system at time $t_{n,4}$;
\item the farthest point it reaches in the system is $d_n$, i.e. $d_n=\max_t I_n(t)$;
\item it enters the ``tip'' at time $t_{n,2}$, i.e. this is the smallest $t$ such that $I_n(t)>L_{tip}$;
\item it leaves the ``tip'' at time $t_{n,3}$, i.e. this is the largest $t$ such that $I_n(t)>L_{tip}$;
\end{itemize}
where $L_{tip}$ refers to the farthest site away from the plus end at the tip, typically we set $L_{tip}\approx 90\% N$ for simulations. These times are illustrated in Figure~\ref{fig_eff_time}~(left panel) with a comparison of endosome motility {\em in vivo} in Figure~\ref{fig_eff_time}~(right panel). If the particle does not reach the tip then we have only $t_{n,1}<t_{n,4}$ defined, while if $d_n\geq L_{tip}$ then we have $t_{n,1}<t_{n,2}<t_{n,3}<t_{n,4}$ defined. Based on these notations, we define the {\em particle arrival efficiency} to be \footnote{$T_0$ is assumed to be large enough that any transients have decayed, i.e., $t_{1,1}\geq T_0$.}
$$
E_1=\lim_{T\rightarrow \infty}E_1(T)=\lim_{T\rightarrow\infty}\frac{\#\{n~:~d_n\geq L_{tip}, ~t_{n,4}<T\}}{\#\{n~:~T_0<t_{n,1}<t_{n,4}<T\}}.
$$
Note that $0\leq E_1\leq 1$ is a dimensionless measurement of efficiency that is close to zero if few of particles arrive at the tip before leaving the system while it is close to one if most of particles arrive at the tip before leaving the system. The quantity $E_1$ is independent of how long it takes to arrive at the tip or leave the system. Meanwhile, we define the {\em average tip delay} to be
$$
E_2=\lim_{T\rightarrow \infty}E_2(T)=\lim_{T\rightarrow \infty} \frac{\sum_{\{n~:~||d_n||\geq L_{tip},t_{n,4}<T\}} (t_{n,3}-t_{n,2})}{\#\{n~:~d_n \geq L_{tip}, ~t_{n,4}<T\}}.
$$
The average tip delay $E_2$ has unit of time, and measures the average time spent at the tip for all particles that get there: larger values indicate that particles are trapped at the tip for a long time while small values indicate that the particles wait only a short time before leaving the tip. Note that the quantity $E_2$ is independent of $E_1$, the proportion of particles that arrive at the tip. Note also that if we assume homogeneous particle-type change rates then $E_2\geq 1/p_{+-}$.

\begin{figure}
\begin{minipage}{0.5\textwidth}
\centering
\includegraphics[scale=0.45]{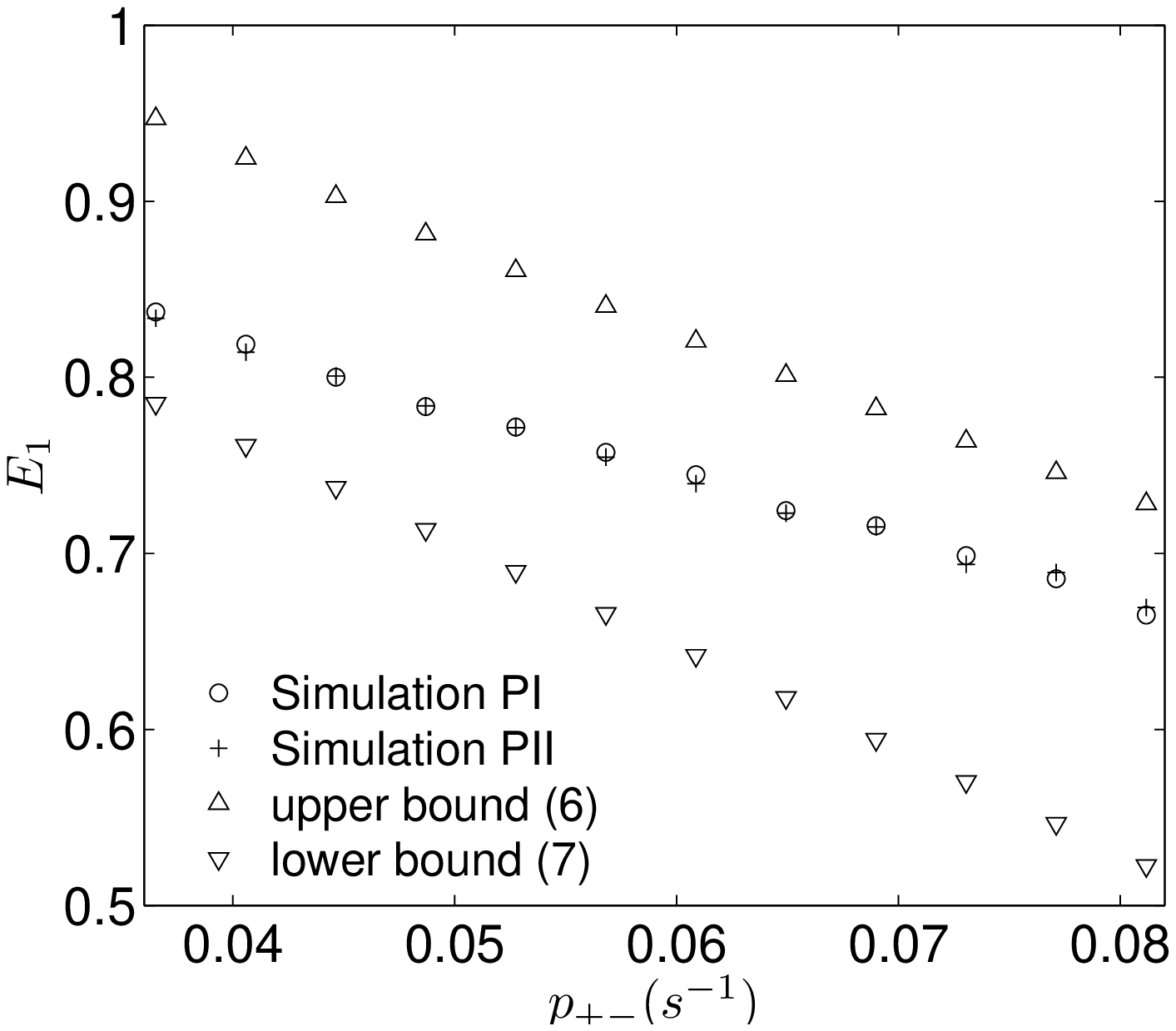}\\
(a)
\end{minipage}
\begin{minipage}{0.5\textwidth}
\centering
\includegraphics[scale=0.45]{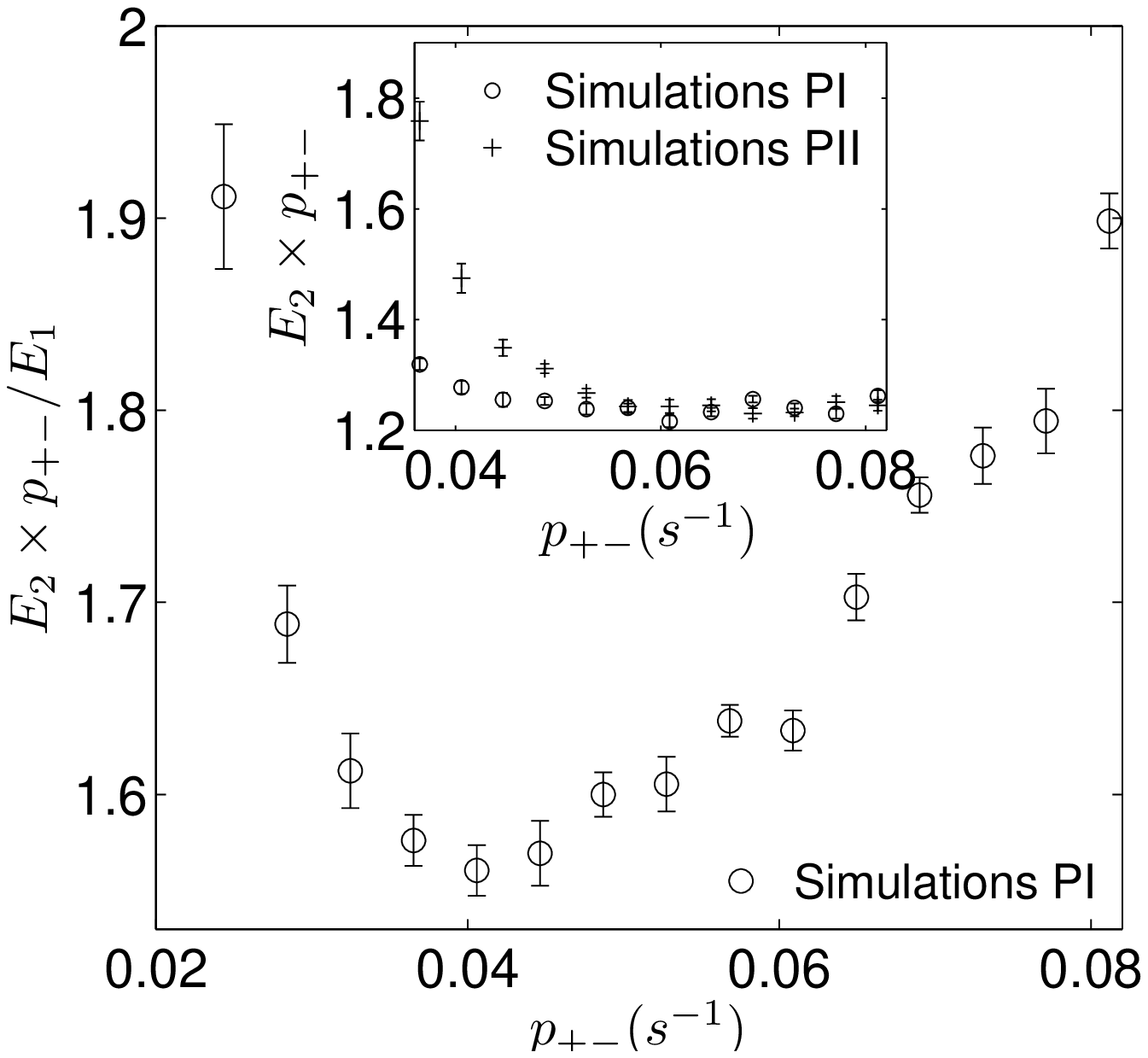}\\
(b)
\end{minipage}
\caption{
Illustration of $E_1$ and $E_2$ when varying $p_{+-}$. (a) shows that the arrival efficiency $E_1$ in both lane-change protocols PI and PII behaves similarly as varying $p_{+-}$ with approximate upper bound from (\ref{eq_E1up}) and lower bound from (\ref{eq_E1low}). (b) shows the balance between delay ratio and arrival efficiency quantified as $E_2\times p_{+-}/E_1$ for protocol PI; the inset shows the comparison between two protocols PI and PII of the delay ratio $E_2\times p_{+-}$; errorbars are indicated. Simulations are done by averaging over 10 runs using the Gillespie algorithm in a time interval $[T_0,T]=[500s, 3000s]$; other rates are default as in Table~\ref{tab_thirteen} and boundary conditions~(\ref{eq_boundary}).
}\label{fig_E1}
\end{figure}

One can derive approximate upper and lower bounds of $E_1$ for the lane homogeneous multi-lane model. Considering the density of plus-type particles in dilute situation; see density approximation~(\ref{equ_den_13lane}) in~\ref{sec_homo}, we have
\begin{eqnarray}\label{eq_E1up}
E_1&\leq&\min\{\rho(1)/\rho(0),1\}\\
&=&\min\left\{\exp\left[\frac{Np_{-+}}{p_-+p^{l\rightarrow l+1}_{-,u}+p^{l\rightarrow l-1}_{+,u}}-\frac{Np_{+-}}{p_+
+p^{l\rightarrow l+1}_{+,u}+p^{l\rightarrow l-1}_{+,u}}\right],1\right\}\nonumber
\end{eqnarray}
while considering the probability of first particle-type change occurring at the tip, we have
\begin{equation}\label{eq_E1low}
E_1\geq \left[1-\frac{p_{+-}}{p_+ + p^{l\rightarrow l+1}_{+,u}+
p^{l\rightarrow l-1}_{+,u}}\right]^{L_{tip}}\approx 1-\frac{Np_{+-}}{p_+ + p^{l\rightarrow l+1}_{+,u}+
p^{l\rightarrow l-1}_{+,u}}
\end{equation}
as particles may change directions (type changes) several times before reaching the tip. Figure~\ref{fig_E1}~(a) shows the upper and lower bounds for $E_1$ compared to the simulations. Moreover, from Figure~\ref{fig_E1}~(a) and Table~\ref{tab_eff}, this arrival efficiency is independent on the particle-type change rate.

The average tip delay $E_2$ is clearly dependent on $p_{+-}$ for a homogeneous model as $1/p_{+-}$ is the expected delay of minus-type particles without obstruction such as in the two-lane model recalled in Section~\ref{sec_twolane} (see \cite{Ashwin_etal_2010} for details). We therefore consider the {\em delay ratio} $E_2\times p_{+-}$, i.e., the ratio of the average tip delay to the expected delay if there are no trapped minus-type particles. This ratio is usually no less than 1, and when it is close to 1 this is the most efficient situation in terms of the delay being minimal.  Simulation results in Table~\ref{tab_eff} and the inset to Figure~\ref{fig_E1} show that for both lane-change protocols in the multi-lane model, this delay ratio is larger than 1 or $E_2>1/p_{+-}$. This agrees with results in Table~\ref{tab_trapped} showing the existence of certain proportion of minus-type particles being ``trapped'' at the tip. Comparing Figure~\ref{fig_E1}~(a) and the inset in Figure~\ref{fig_E1}~(b), we consider the dimensionless quantity $E_2\times p_{+-}/E_1$ to be a balance between arrival efficiency and delay ratio. As varying $p_{+-}$ we find a minimum value where $p_{+-}$ is closed to the experiment data of $p_{+-}=0.0406~s^{-1}$; see Figure~\ref{fig_E1}~(b). This offers an implication of {\em in vivo} transport rates.

Comparing the two lane-change protocols PI and PII, protocol PI is more effective in the sense that it has a relatively lower delay ratio $E_2\times p_{+-}$ when varying $\alpha_+$ (see Table~\ref{tab_eff}) or for small $p_{+-}$ (see the inset in Figure~\ref{fig_E1}~(b)). Similar to the case for trapped minus-type fraction and mean tip size, the average tip delay time increases more rapidly with injection rate in protocol PII than the increase in protocol PI. 

\begin{table}
\caption{\label{tab_eff}
Comparison of $E_1,E_2$ between lane-change protocols PI and PII. The arrival efficiency $E_1$ in both protocols is relatively independent of $\alpha_+$ and drops to almost zero when $\alpha_+>\alpha_{c,I}$ for protocols PI ($\alpha_+>\alpha_{c,II}$ for protocol PII). The average tip delay $E_2$ in both protocols are larger than the expected delay $1/p_{+-}\approx 24.6~s^{-1}$ when no minus-type particles are trapped, and the delay time increases more rapidly for protocol PII than for PI when increasing $\alpha_+$. Note that the particle-type change rate $p_{+-}$ is fixed here, meaning that the delay ratio $E_2\times p_{+-}$ is a multiple of $E_2$. Simulations are done as in Figure~\ref{fig_E1} with otherwise default parameters.
}
\begin{indented}
\item[]\begin{tabular}{@{}l|l|llllllll}
\br
& $\alpha_+(s^{-1})$  & 0.8 & 0.9 & 1 & 1.1 & 1.2 & 1.3 & 1.4 & 2\cr
\mr
PI & $E_1$ & 81.9\%& 81.9\% & 81.5\% & 81.9\%  & 81.6\% & 81.6\% & 81.9\%  & 81.7\% \cr
& $E_2$ (s) & 30.71 & 30.87 & 31.1  & 31.1 & 31.8 & 32.6 &  33.5  & 48.8 \cr
\mr
PII & $E_1$ & 81.4\% & 81.8\%& 81.7\% & 81.7\%  & 81.9\% & 81.8\% & 81.5\%&  $\approx$ 0  \cr
& $E_2$ (s) & 31.9 & 33.2 & 34.2  & 37.9 & 44.9 & 48.7 & 61.4  & --\cr
\br
\end{tabular}%\end{array}
\end{indented}
\end{table}

\section{Critical behaviour and pulsing states}
\label{sec_critical}

Results from Section~\ref{sec_tipsize} suggest that the mean tip size does not converge when an injection rate $\alpha_+$ approaches a critical injection rate $\alpha_{c,I}$ for lane-change protocol PI ($\alpha_{c,II}$ for protocol PII) in a finite system; additionally, by simulating the efficiency defined in terms of delay ratio $E_2\times p_{+-}$, the system will be less efficient with either a large total injection rate $\alpha_+$ (see Table~\ref{tab_eff}) or a small particle-type change rate $p_{+-}$ (see Figure~\ref{fig_E1}~(b)). Large mean tip size or inefficiency on $E_2\times p_{+-}$ are expected due to high densities of plus- and minus-type particles, and in the following we discuss the dynamics of pulsing states that exist in this model.

\subsection{Pulsing states in the system}
\label{sec_pulsing}

\begin{figure}
$~t=50000\times h_t$\\
\centerline{\epsfig{file=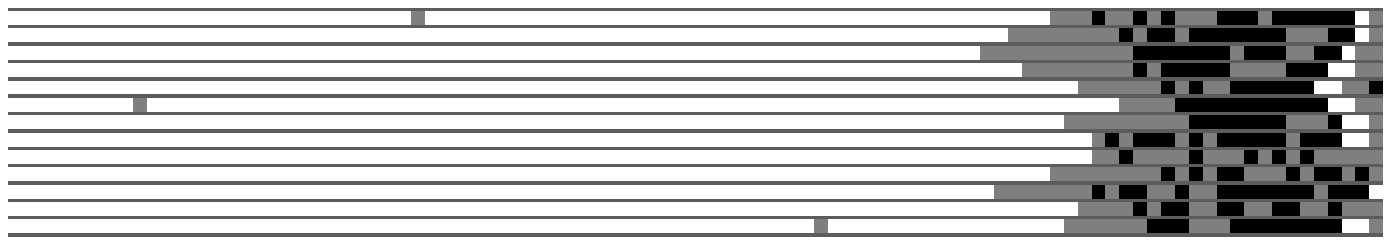,width=10cm}}

$~t=70000\times h_t$\\
\centerline{\epsfig{file=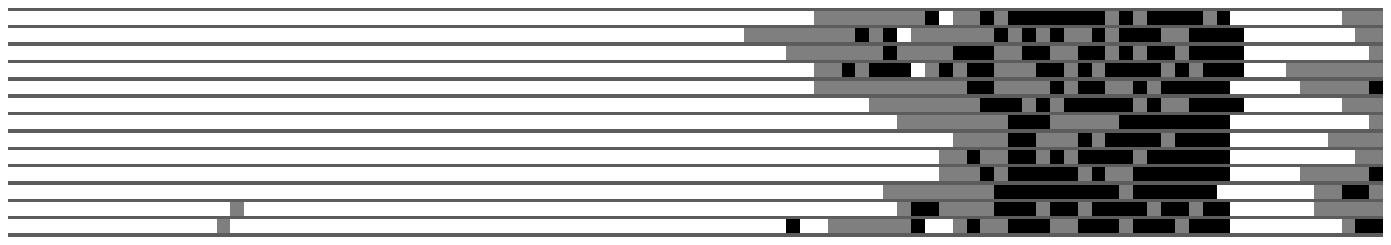,width=10cm}}

$t=140000\times h_t$\\
\centerline{\epsfig{file=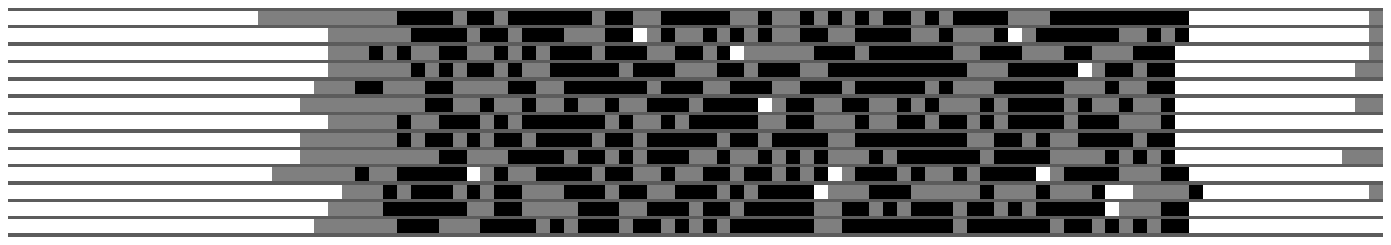,width=10cm}}

$t=160000 \times h_t$\\
\centerline{\epsfig{file=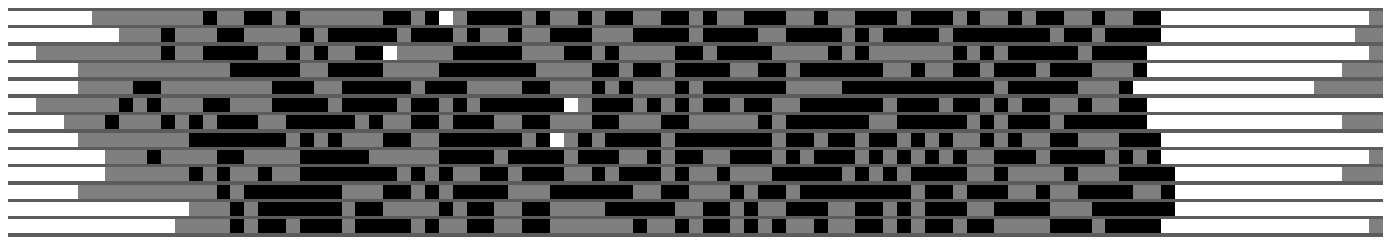,width=10cm}}

$t=300000 \times h_t$\\
\centerline{\epsfig{file=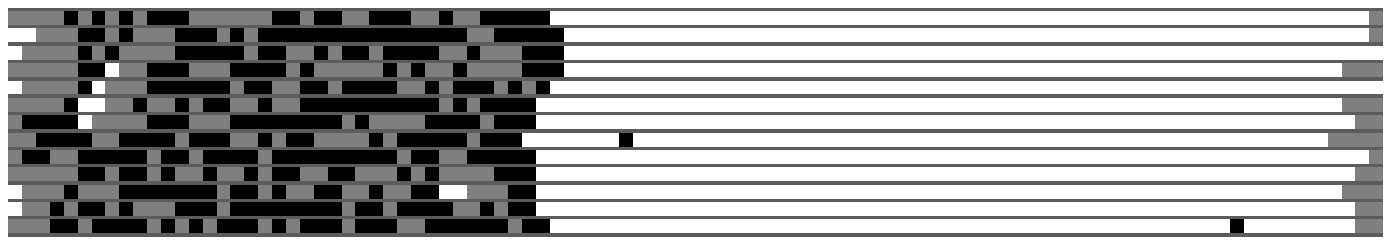,width=10cm}}

$t=340000 \times h_t$\\
\centerline{\epsfig{file=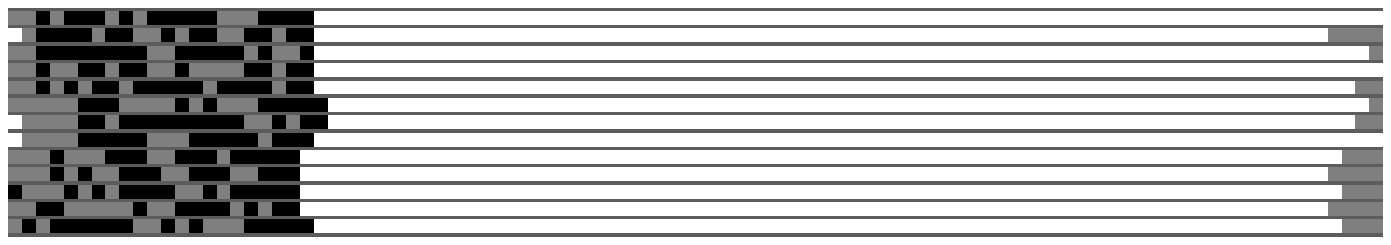,width=10cm}}
\caption{
Time progress of a pulsing state for the thirteen lane model of size $N=100$, $\alpha_+=4.0~s^{-1}$ and no flux on the plus-end; otherwise parameters as in Table~\ref{tab_thirteen} and boundary conditions as in (\ref{eq_boundary}). The sequence of filling (top three frames) and emptying (bottom three frames) repeats approximately periodically. Observe that the pulse of high-density mixed particles propagates slowly to the left via diffusion of vacancies through the pulse. Grey indicates plus-type particles (that move to the right) while black indicates minus-type particles (that move to the left). The white regions indicate vacancies.
}\label{fig_time_pulse}
\end{figure}

For injection rate $\alpha_+$ smaller than the critical value $\alpha_c$ discussed in Section~\ref{sec_tipsize}, the system converges to a stationary state ({\em  shock state}) with an accumulation of certain size at the tip, and this tip size increases as $\alpha_+$ approaching $\alpha_c$. For $\alpha_+>\alpha_c$ and otherwise default parameters, a new type of behaviour that we call a {\em pulsing state} can appear. These are characterized by the behaviour in Figure~\ref{fig_time_pulse}. For finite systems, near $\alpha_c$ the accumulation at the tip may show large irregular oscillations including a propagating region or {\em pulse} of high density that moves away from the tip before dispersing.

\begin{figure}
\begin{minipage}{0.5\textwidth}
\centering
\includegraphics[scale=0.45]{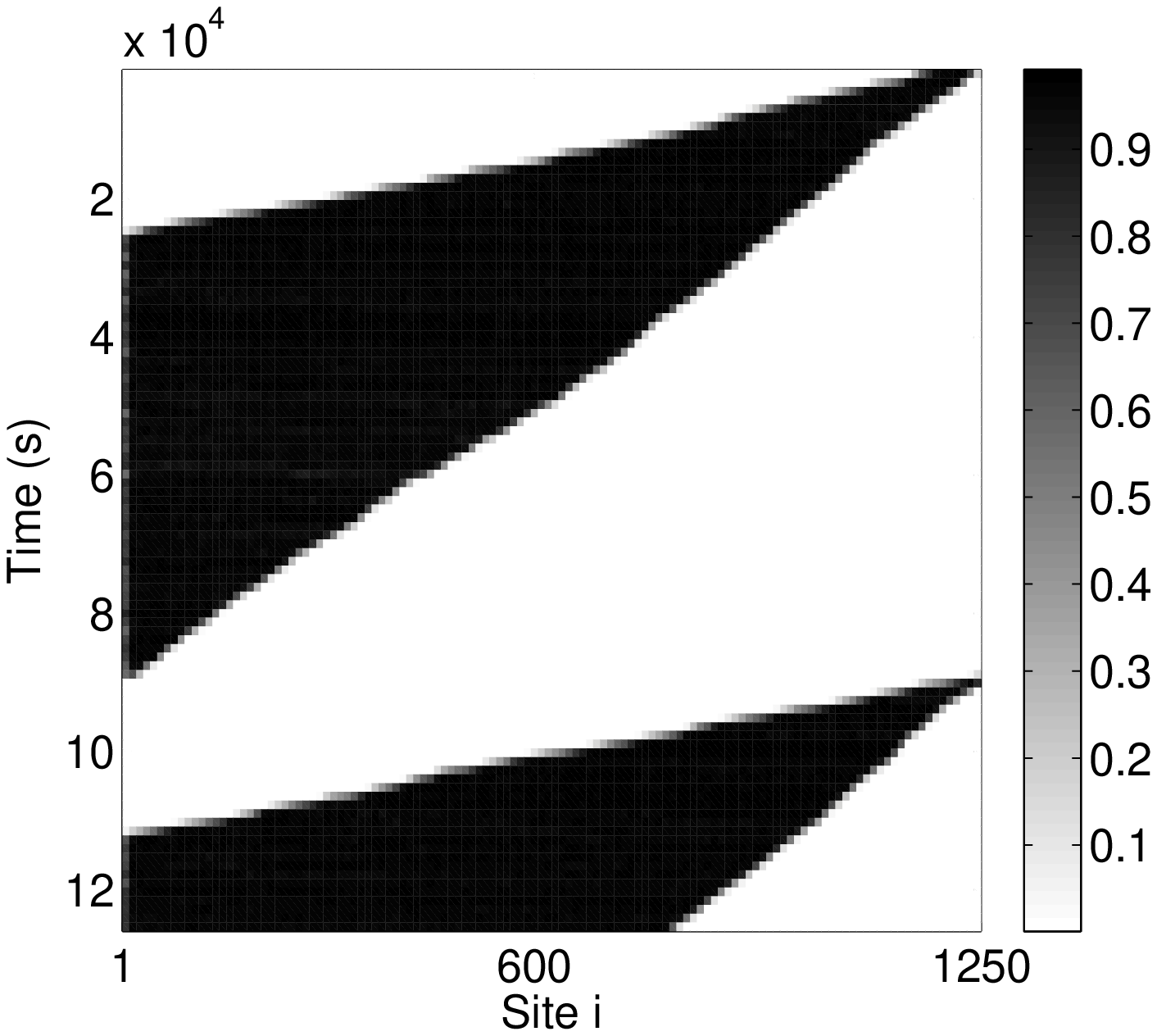}\\
(a)
\end{minipage}
\begin{minipage}{0.5\textwidth}
\centering
\includegraphics[scale=0.45]{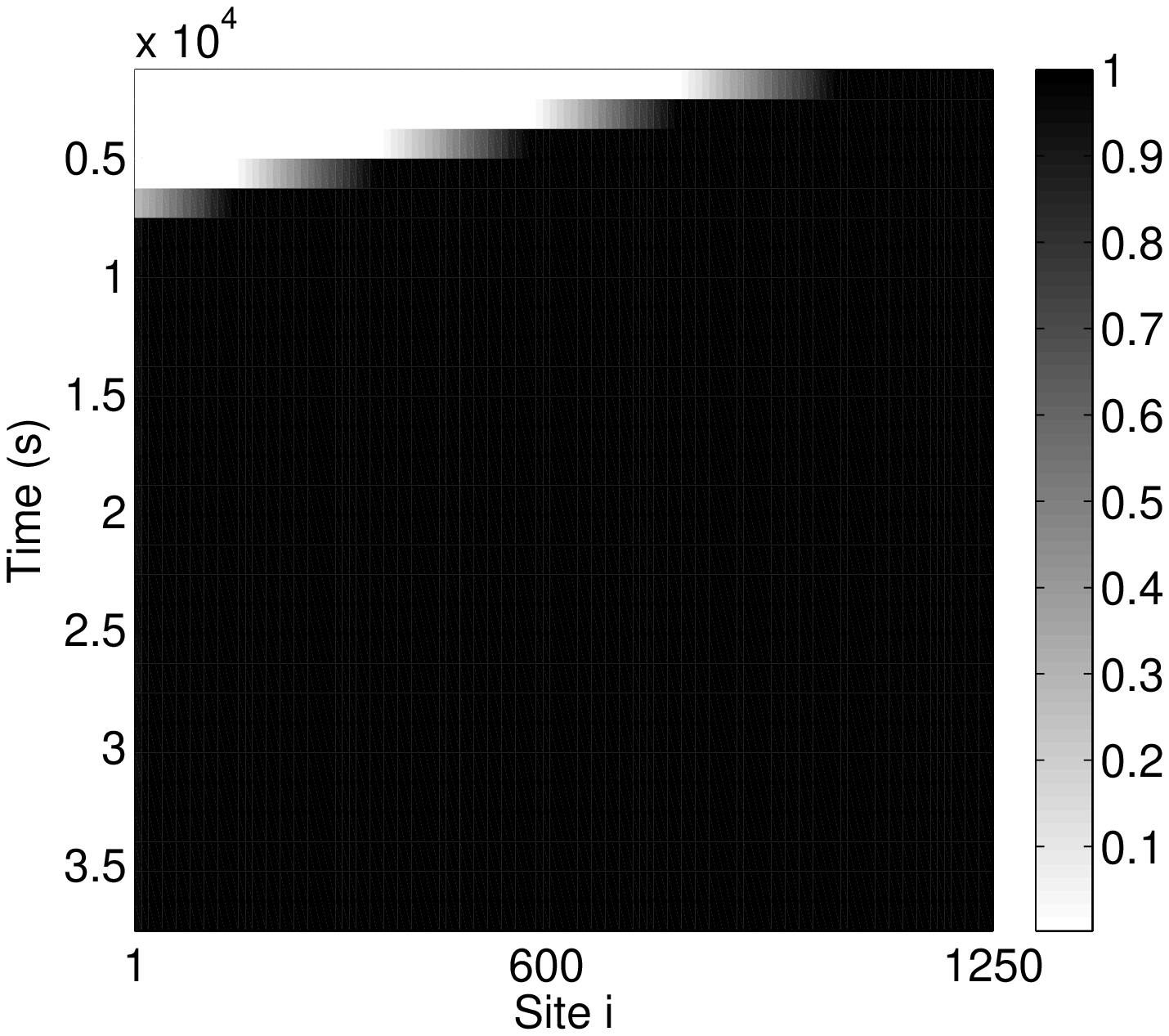}\\
(b)
\end{minipage}
\caption{
(a) shows a pulsing state for $p_{-+}=0.05~s^{-1}$ and $\alpha_+= 3~s^{-1}$; (b) shows a filled state for $p_{-+}=0.14~s^{-1}$ and $\alpha_+=4~s^{-1}$ and this approaches a homogeneous high-density mixed equilibrium state. The other rates are default as in Table~\ref{tab_thirteen} and boundary conditions in (\ref{eq_boundary}).
}\label{fig_turn}
\end{figure}

In a {\em pulsing state}, the whole system density undertakes large approximately periodic oscillations. These can be split into two parts:
\begin{itemize}
\item
A {\em filling phase} during which the density at the minus end is low, and a growing high density {\em pulse} of mixed particles moves steadily towards the minus end. During this phase there is a net flux into the system.
\item
An {\em emptying phase} during which the density at the minus-end is high, and the pulse propagates out of the system. During this phase there is a net flux out of the system.
\end{itemize}
For most of the cycle there is a low density region between the tip and the pulse. The pulsing occurs through an alternation between these phases as fronts (evident in Figure~\ref{fig_time_pulse}) separating low and high density regions move through the domain. At other parameters we find the system converges to a {\em filled state} for $\alpha_+$ larger than $\alpha_c$. Figure~\ref{fig_turn} shows examples of the pulsing and filled state. In the filled state, once the filling phase is complete the system remains approximately full.

\begin{figure}
\centerline{\epsfig{file=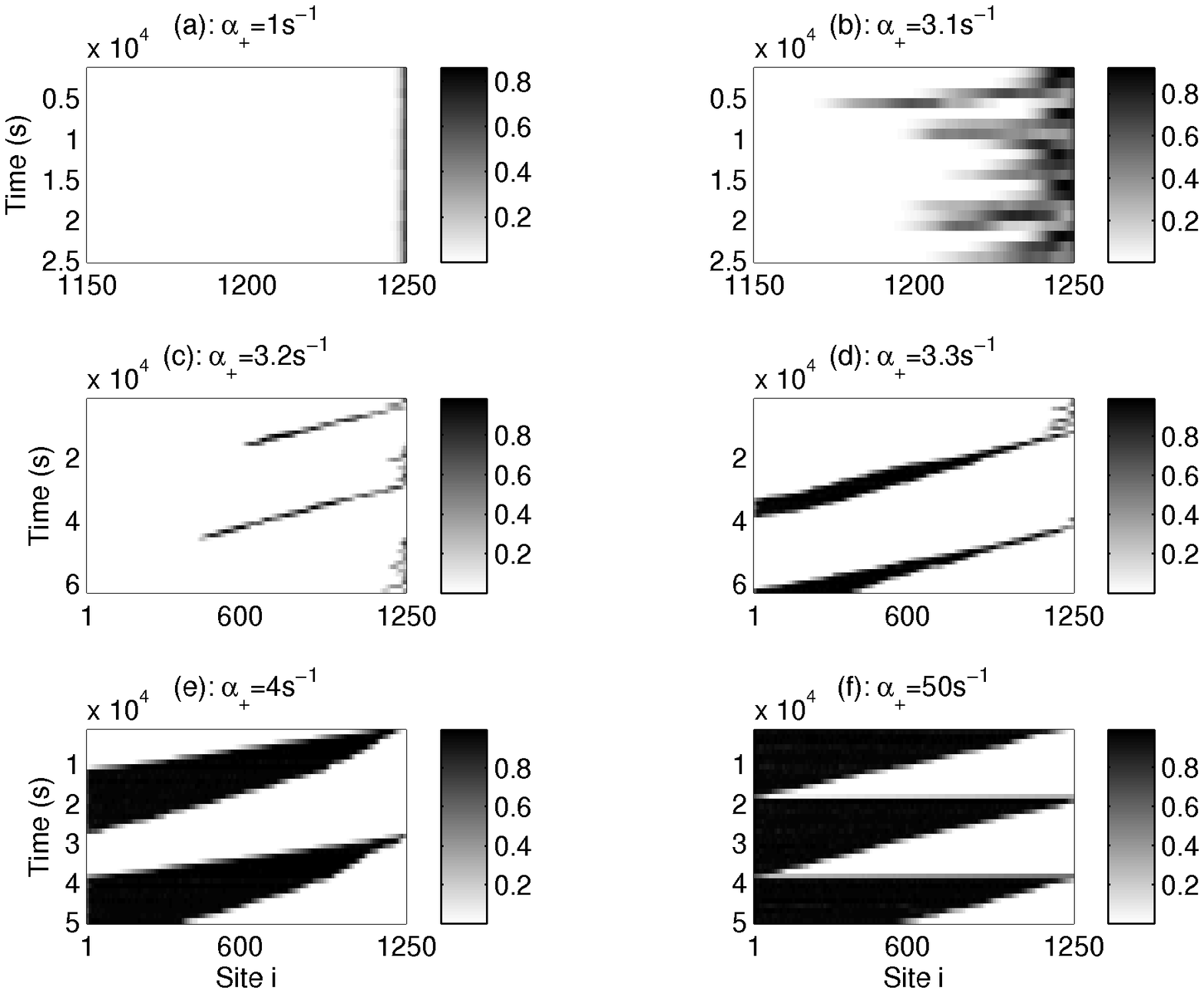,width=16cm}}
\caption{
Each horizontal line represent a time-average of the occupancy $\sum_{l=1}^{M}(\tau^l_{-,i}+\tau^l_{+,i})/M$ over blocks of length $1260~s$ in the homogeneous thirteen-lane model ($M=13$) with injection rate $\alpha_+$ indicated (other rates as default). In (a), the model with a small total injection rate reaches its stationary state with a small number of particles accumulated in the tip. (b)-(d) show the density-time courses for injection rates near the critical value.  In (d)-(f), pulsing states appear where the accumulation moves away from the plus end and simultaneously grows.
}\label{fig_density_pulse}
\end{figure}

To better understand the behaviour of system in a pulsing state, Figure~\ref{fig_density_pulse} shows successive horizontal lines as the evolution of the medium time average of the density $\sum_{i=1}^M(\tau_{-,i}+\tau_{+,i})/M$ (over successive blocks of $1260~s$) from a time-series and gradual changes of the local density. Figure~\ref{fig_density_pulse}~(a,b) shows rapid convergence to a stable accumulation at the tip, though (b) displays more fluctuations. Examples of pulsing states are illustrated in Figure~\ref{fig_density_pulse}~(d,e,f) for increasing values of $\alpha_+$; these exhibit long approximately periodic oscillations. Figure~\ref{fig_density_pulse}~(c) shows an intermediate situation where the tip accumulation shows irregular large amplitude oscillations. The main behaviour in a high-density mixed region will be changes of particle types and therefore we expect the ratio of densities between minus- and plus-type particles to reach the equilibrium ratio given by the type-change rates, i.e.
$$
\tilde{\rho}=\frac{p_{-+}}{p_{-+}+p_{+-}}\kappa,~~\tilde{\sigma}=\frac{p_{+-}}{p_{-+}+p_{+-}}\kappa
$$
where $\kappa=\tilde{\rho}+\tilde{\sigma}\approx 1$ represents the total local density. 

For both lane-change protocols PI and PII we find pulsing states for $\alpha_+>\alpha_c$ and otherwise default parameters - we interpret this as an existence of a negative net flux for the high-density mixed phase.  Note that densities $\tilde{\rho}$ and $\tilde{\sigma}$ are primarily governed by the particle-type change rates $p_{+-}$ and $p_{-+}$, if we increase $p_{-+}$, then the net flux may change to be positive and the pulsing state will be replaced by approach to a high-density uniform filled state. For example, Figure~\ref{fig_turn}~(a) shows the pulsing state for $p_{-+}=0.05~s^{-1}$ with $\alpha_+=3~s^{-1}$ while Figure~\ref{fig_turn}~(b) shows a filled state for $p_{-+}=0.14~s^{-1}$ and $\alpha_+=4~s^{-1}$ (other parameters are the default values). For the latter case, we infer that an initial pulse with $j>0$ will saturate to $\tilde{\rho}+\tilde{\sigma}=1$ and then $j\rightarrow 0$ as the number of vacancies in the high density state goes to zero. It will be an interesting challenge to understand the statistical properties of these high-density mixed states and, for example, to predict $j$ from the parameters of the system, as there might be more than one particle compete for a common empty site.

\begin{figure}
\begin{minipage}{0.5\textwidth}
\centering
\includegraphics[scale=0.45]{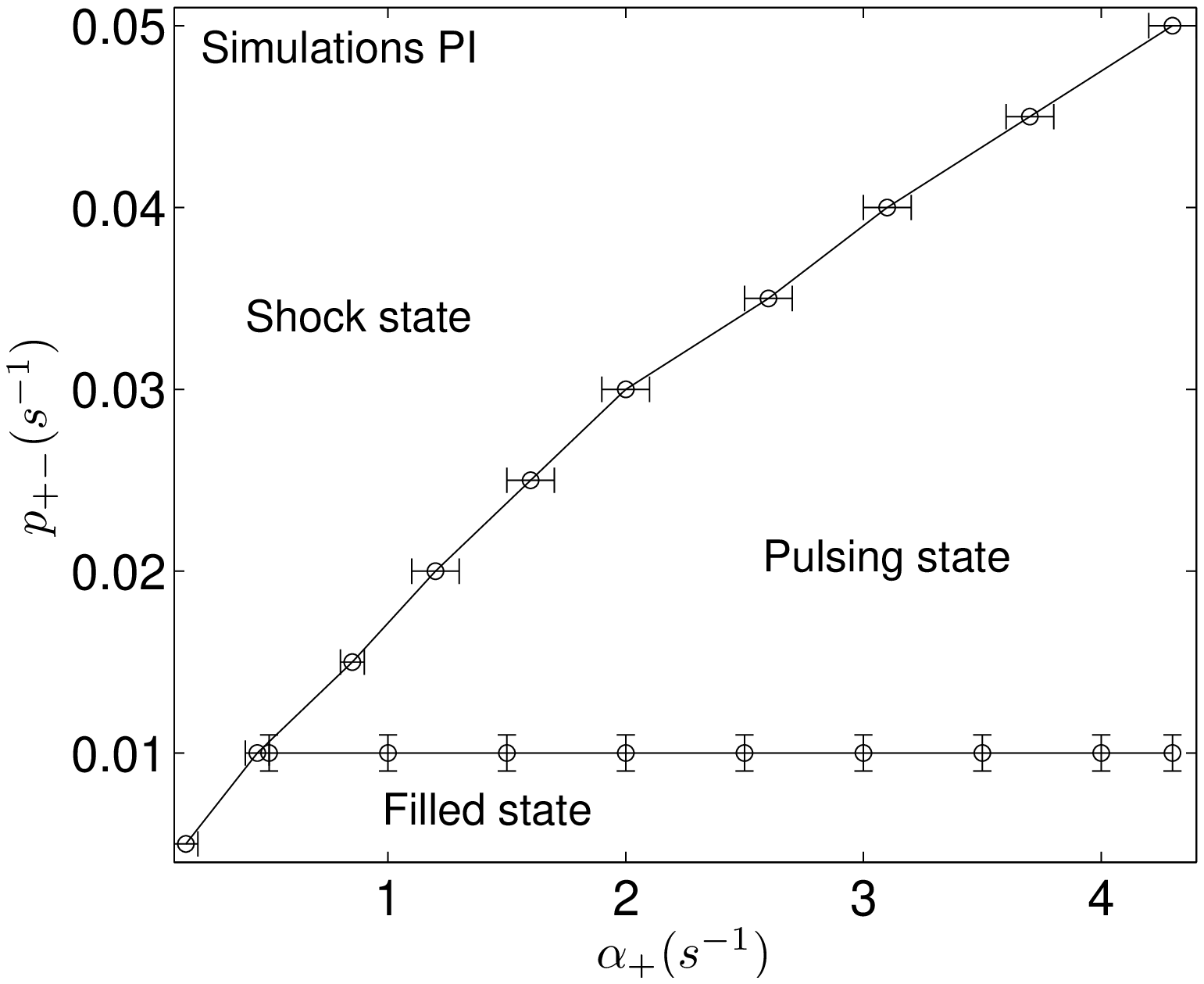}\\
\end{minipage}
\begin{minipage}{0.5\textwidth}
\centering
\includegraphics[scale=0.45]{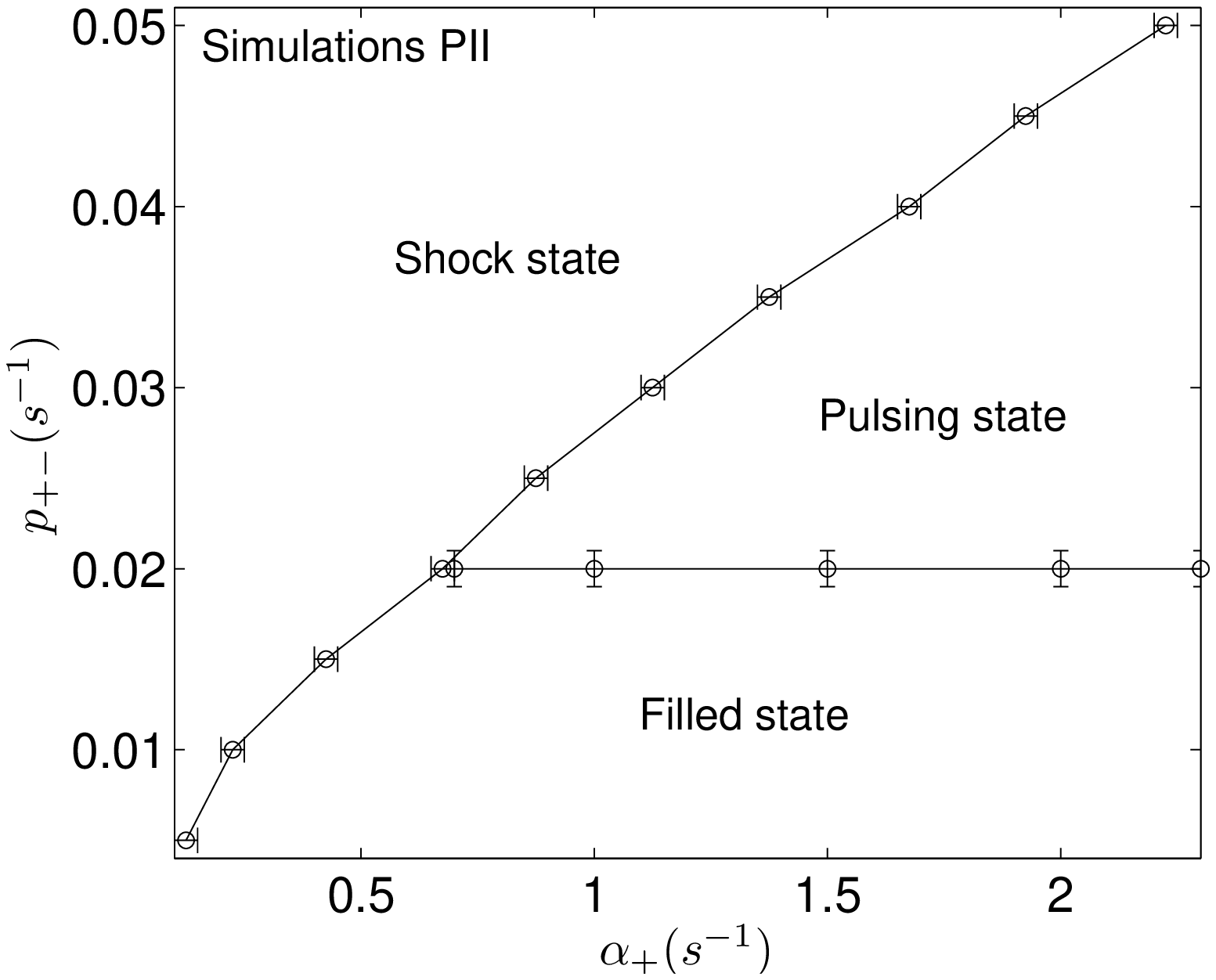}\\
\end{minipage}
\caption{
Numerically obtained phase diagrams for lane change protocol PI (left panel) and lane change protocol PII (right panel). Other default parameters are as in Table~\ref{tab_thirteen} and boundary conditions as in (\ref{eq_boundary}). Bars indicate uncertainty of the borders between different states.
}\label{fig_phase}
\end{figure}

\begin{figure}[ht]
\centerline{\epsfig{file=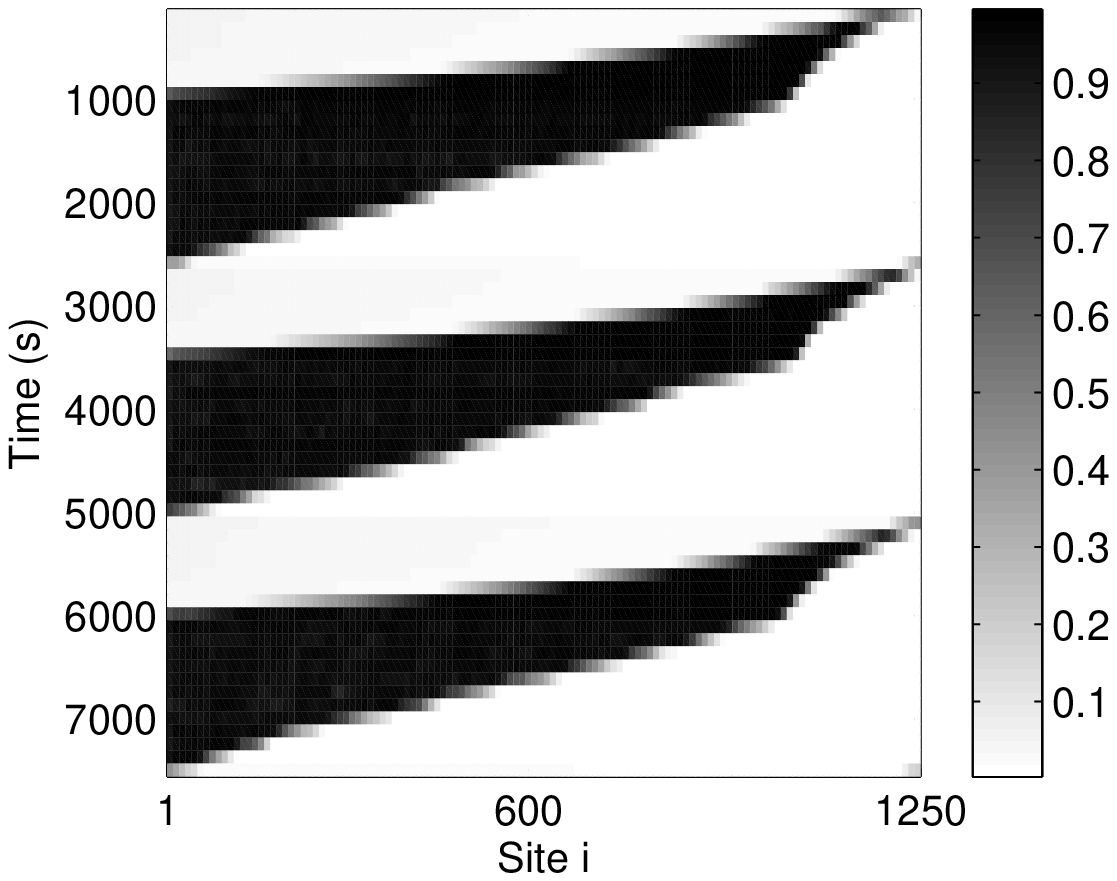,width=7cm}}
\caption{
A pulsing state for $p_{-+}=0.2~s^{-1}$, $p_{+-}=0.4~s^{-1}$ and $\alpha_+= 80~s^{-1}$; other parameters are default as in Table~\ref{tab_thirteen} and boundary conditions in (\ref{eq_boundary}).
}\label{fig_pulshihigh}
\end{figure}

Numerical simulations show that this pulsing state is robust to changes in parameters -- Figure~\ref{fig_phase} shows phase diagrams for lane-change protocols PI and PII on varying $\alpha_+$ and $p_{+-}$ with other default parameters. The particle-type change rates we usually use obeys $\frac{p_{+-}}{p_+}, \frac{p_{-+}}{p_-} < \frac{1}{N}$ which means that a typical switching of plus-type particle occurs at the tip (see $E_1$ in Table~\ref{tab_eff}), this novel pulsing state also appears when $\frac{p_{+-}}{p_+}, \frac{p_{-+}}{p_-} >\frac{1}{N}$; see Figure~\ref{fig_pulshihigh}. A necessary condition for a pulsing state we suggest is that $\alpha_+$ is beyond a critical value (i.e., not in a shock state) and there is a net negative flux in the high-density mixed region.

\subsection{System size effects}

Although the previous section focuses on a fixed system size $N$, the critical transition to pulsing states is invariant of $N$ as $N\rightarrow \infty$ under certain scalings of parameters for the multi-lane model with boundary conditions~(\ref{eq_boundary}). For example, if we scale the forward motion rates, keeping $p_{\pm}/N$ constant and use the lane change rates when blocked $p_{+}/2$ as default whilst keeping the particle-type change rates constant, we find no significant variation of mean tip size with $N$ as illustrated in Table~\ref{tab_sizeeff}. This agrees with the approximation~(\ref{eq_ntip}) from the two-lane model. For injection rates beyond a critical value and the same scaling we find pulsing states for arbitrarily large $N$.

\begin{table}
\caption{\label{tab_sizeeff}
The tip size when varying the system size $N$ for parameters in Table~\ref{tab_thirteen} and boundary conditions~(\ref{eq_boundary}) except keeping $p_+/N=0.17$, $p_-/N=0.1631$ and $p_{-,b}=p_+/2$. There values are chosen to be consistent with rates in Table~\ref{tab_thirteen} when $N=1250$. Mean and standard error of mean are shown.
}
\begin{indented}
\item[]\begin{tabular}{@{}c|l|l|l|l|l}
\br
$N$ & 400  &800 & 1200& 1600 &2000 \cr %& 2200\cr
\mr

$<n_{tip}>\pm$ sem & 26.7$\pm$0.278 & 26.9$\pm$ 2.7& 26.8$\pm$ 2.97 & 26.9$\pm$2.9 & 26.7$\pm$2.9\cr %& 26.7$\pm$2.99\cr
\br
\end{tabular}%\end{array}
\end{indented}
\end{table}

\subsection{A simple model with a critical injection rate}
\label{sec_singlelane}

To better understand the appearance of a critical injection rate $\alpha_c$ that leads to a transition in the behaviour of the tip, we introduce a simpler model that computably predicts nonlinear behaviour of the mean tip size with injection rate $\alpha_+$, and a singularity of the mean tip size at finite injection rate.

Consider a single lane first-in, last-out queue of plus-type and minus-type particles. Plus-type particles are assumed to arrive at the left end of the queue at a rate $\alpha$ per time-step. Every time-step, we assume that all particles independently to be minus-type with probability $p$ and plus-type with probability $1-p$. A number of particles is assumed to leave the queue whenever the left-most particle is of minus-type, in which case all of the adjacent minus-type particles are assumed (instantaneously) to leave the queue. This results in a steady growth and intermittent loss of particles. At a given time-step $t$, if we assume the queue has size $n(t)$ then after random changes in type, the number of particles that leave the queue can be expressed as
$$
\lambda(n)=\sum_{m=1}^{n} mp^m(1-p) + n p^{n}= p\frac{1-p^n}{1-p}.
$$
Hence the mean rate of net growth of the queue will be given by $\alpha_+ -\lambda(n)$. This will give zero net growth when $\alpha_+ - \lambda(n)=0$. Therefore, with injection rate $\alpha_+$, the mean size of the queue is
$$
\left<n(\alpha_+)\right>=\frac{ \ln\left(1-\alpha\frac{1-p}{p}\right)}{\ln p},
$$
which predicts a critical injection rate $\alpha_c=\frac{p}{1-p}$. For $\alpha_+<\alpha_c$, $\left<n(\alpha_+)\right>$ grows monotonically and nonlinearly with $\lim_{\alpha_+\rightarrow\alpha_c-}\left<n\right>(\alpha)=\infty$. These are clear analogies with the observations of mean tip size in the thirteen-lane model in Figure~\ref{fig_tipsize}~(a). Indeed, one can fit the data in Figure~\ref{fig_tipsize}~(a) to a logarithmic function but the fit is inferior to the rational function discussed there, suggesting that the model above is too simple for an accurate quantitative explanation. 
This simple model reproduces the nonlinear increase for $\alpha_+<\alpha_c$ in Figure~\ref{fig_tipsize}~(a) and the singular behaviour at the critical value $\alpha_c$. For $\alpha_+>\alpha_c$ it predicts unbounded growth of the queue, analogous to convergence of the system to a ``filled state''.

\section{Discussion}
\label{sec_shortdisc}

Motor-driven bidirectional transport of vesicles and organelles is vital for the organization and function of eukaryotic cells and intracellular motility serves various cellular processes \cite{WelGro_2008}. The basic function of bidirectional vesicle transport is to deliver cargo over distances, thereby modifying gradients and ensuring communication between different regions of the cell.

By considering the process in a number of stages, starting with cargo uptake, followed by transport along the fibres of the cytoskeleton, and finally ending with cargo off-loading, the site of unloading is often also a region of cargo uptake. Transport back on the cytoskeletal track therefore not only recycles the transport vesicles, but also serves for long-distance delivery to other regions of the cell. In a fungal model system, it has recently been shown that dynein are concentrated at MT plus-ends to prevent organelles falling off the track and this concentration of dynein is done by an active retention mechanism (based on controlled protein-protein interaction) and stochastic motility of motors \cite{Schuster_etal_2010,Schuster_etal_2011}. 

An accumulation of motors at MT plus-ends can be seen as an inefficiency, as dynein motors are supposed to transport particles rather than wait at the MT end. Thus, the cell needs to find a compromise between (a) ensuring that organelles are captured at MT ends, which requires dynein accumulation at the plus-ends \cite{Schuster_etal_2010} and (b) keeping dynein moving along MTs to deliver the cargo to minus-ends.
As discussed in Section~\ref{sec_efficiency}, an interesting result of this study is that the parameters from \cite{Schuster_etal_2010} in a {\em Ustilago maydis} system (and $p_{+-}$ in particular) do address this compromise between (a) and (b). Moreover on comparing different lane-change protocols (that is, either dynein changes lanes or both kinesin-1 and dynein change lanes), we find that it is necessary for only dynein to change lanes in order to give biologically realistic system states over a larger range of possible fluxes; see Figure~\ref{fig_tipsize}~(a). 

The maximum unidirectional flux in the thirteen-lane model can occur, by analogy with the so-called {\em maximal current state} in the ASEP model \cite{Derrida_etal_1993, BlytheEvans_2007}, when $\alpha_+\approx~13\times~p_+/2\approx 1300~s^{-1}$, assuming appropriate boundary conditions and only one particle type. Note that the maximal flux with bidirectional transport in the thirteen-lane model is about the half of that in unidirectional transport. However, the existence of a critical injection rate $\alpha_c\ll 1300/2~s^{-1}$ (see Figure~\ref{fig_tipsize}) in the half-closed homogeneous thirteen-lane model places a much lower bound on the maximum flux. The experimentally measured $\alpha_+\approx 1.06~s^{-1}$ in Table~\ref{tab_thirteen} is of the same order as $\alpha_c$. To achieve bidirectional flux that is half of the maximum unidirectional flux (about $600~s^{-1}$) on a MT, much more organized transport with specific control of lane-change to segregate different particles into different lanes (as in Table~\ref{tab_twolane}) is therefore necessary.

For the half-closed system, we have found for $\alpha_+>\alpha_c$ some novel ``pulsing states'' of the system. It will be very interesting to further explore the region of existence of these states. There may exist other new phases for other boundary conditions and/or transition rates in the multi-lane model. An open question includes whether one can find a better understanding of the dynamics of pulsing states and/or whether there are analytic approximations that confirm the phase diagram approximated numerically in Figure~\ref{fig_phase}.

The model can be generalized to incorporate a number of features that may be important for MT transport. This includes (a) additional species of motor with different transport properties, such as the several types of kinesin known to exist {\em in vivo} \cite{Val_2003, Wedlich2002}; additional motors can be expected to increase collisions along the MT length and cause high density of motors that might affect the transport efficiencies described in Secton~\ref{sec_efficiency}. (b) The detachment and reattachment of motors from the MT \cite{EbbSan09} may be important in other systems; the size of any accumulation will clearly be affected by this. (c) Non-trivial geometry of MT bundles needs consideration as motors may jump between different MTs; this might be another possibility to avoid collision between counter-moving organelle. (d) A more realistic motor and cargo size will influence the ability of lane changes to overcome blockages. Finally (e) static obstructions such as MAPs (microtubule associated proteins) will influence the behaviour of motors on the MT by increasing more potential blockages
\cite{Dreblow10,Seitz02,Chai_etal_2009,Telley09}; a high concentration of these will clearly influence the transport efficiency. At present, quantitative data for most of these processes are not available for any living cell. With each additional process one can gain quantitatively more precise mathematical representation of the full MT-mediated transport within a cell. It remains to be seen whether new qualitative effects and better understanding will result from more accurate models.

\ack{}
We thank Martin Schuster for acquiring the images that were used to generate the kymograph shown in Figure~\ref{fig_eff_time}~(right panel), and the referees for their insightful comments.

%}
%\smallskip

\newpage

\appendix

\section{Mean field approximation for the multi-lane model}

\label{app_meanfield}

We use a mean field approximation to give an equation for evolution of the densities $\rho^l_i$ and $\sigma^l_i$ for the multi-lane model with lane-homogeneous rates (but not necessarily boundary conditions). Examining the balance of incoming and outgoing particles to a given site, the mean field approximation gives the following, where on the right hand side we write $\rho_i$ to mean $\rho^l_i$, etc.
\begin{eqnarray*}
\frac{d\rho^l_i}{dt}
&=&
p_+\rho_{i-1}(1-\rho_i-\sigma_i)+p_{-+}\sigma_i-p_+\rho_i(1-\rho_{i+1}-\sigma_{i+1})-p_{+-}\rho_i\\
&+&p^{\uparrow}_{+,u}\rho^{l-1}_{i-1}(1-\rho^{l-1}_i-\sigma^{l-1}_i)(1-\rho_i-\sigma_i)\\
&+&p^{\downarrow}_{+,u}\rho^{l+1}_{i-1}(1-\rho^{l+1}_i-\sigma^{l+1}_i)(1-\rho_i-\sigma_i)\\
&+& \left(p^{\uparrow}_{+,b}\rho_{i-1}^{l-1}(\rho^{l-1}_i+\sigma^{l-1}_i)+p^{\downarrow}_{+,b}\rho_{i-1}^{l+1}(\rho^{l+1}_i+\sigma^{l+1}_i)\right)(1-\rho_i-\sigma_i)\\
&-&\left(p^{\uparrow}_{+,u}(1-\rho^{l-1}_{i+1}-\sigma^{l-1}_{i+1})+p^{\downarrow}_{+,u} (1-\rho^{l+1}_{i+1}-\sigma^{l+1}_{i+1})\right) \rho_{i}(1-\rho_{i+1}-\sigma_{i+1})\\
&-&\left(p^{\uparrow}_{+,b}(1-\rho^{l-1}_{i+1}-\sigma^{l-1}_{i+1})+p^{\downarrow}_{+,b}(1-\rho^{l-1}_{i+1}-\sigma^{l-1}_{i-1})\right)\rho_{i}(\rho_{i+1}+\sigma_{i+1})\\
\end{eqnarray*}
There is a similar expression for $\frac{d\sigma^l}{dt}$, but for reasons of space we do not give this here. As the general mean field equations are not easy to solve, we consider below two special cases of these mean field equations.

\subsection{Dilute lane-inhomogeneous densities}
\label{sec_inhom}

Let us assume that
\begin{itemize}
\item $\rho_i^l(t)=\rho^l(x,t)$ with $x=i\delta$ (where $\delta=1/N=h_s/L$ is small) and the spatial dimension is parametrized by $x\in[0,1]$, 
\item $\rho$ and $\sigma$ are small (i.e. dilute limit);
\item the lane changes rates are lane homogeneous and symmetric (i.e., $p_{\pm,u}=p^{\downarrow(\uparrow)}_{\pm,u}$);
\end{itemize}
On expanding $\rho^l_{i+1}=\rho^l(x)+\delta \pard{\rho^l}{x}+O(\delta^2)$ and discarding any terms that are quadratic in $\delta$, the mean field equations above simplify to
\begin{eqnarray*}
\pard{\rho^l}{t} &=& -\delta p_+ \pard{\rho^l}{x} + p_{+,u} \left(\rho^{l+1}+\rho^{l-1}-2\rho^{l}-\delta\pard{\rho^{l-1}}{x}-\delta\pard{\rho^{l+1}}{x}\right)\\
&&+p_{-+}\sigma^l-p_{+-}\rho^l\\
\pard{\sigma^l}{t} &=& \delta p_- \pard{\sigma^l}{x} + p_{-,u} \left(\sigma^{l+1}+\sigma^{l-1}-2\sigma^{l}+\delta\pard{\sigma^{l-1}}{x}+\delta\pard{\sigma^{l+1}}{x}\right)
\\&&+p_{+-}\rho^l-p_{-+}\sigma^l
\end{eqnarray*}
which can be used to characterize the combination of bidirectional transport ($p_{\pm}$), change in direction ($p_{+-}$ and $p_{-+}$) and cross-lane diffusion ($p_{\pm,u}$) on the density. Considering a region of the domain where the dilute approximation holds, the stationary state distribution will therefore satisfy
\begin{eqnarray*}%\label{equ_dilute}
\delta p_+ \derv{\rho^l}{x}&=& p_{+,u} \left(\rho^{l+1}+\rho^{l-1}-2\rho^{l}-\delta\frac{d\rho^{l-1}}{dx}-\delta\frac{d\rho^{l+1}}{dx}\right)+p_{-+}\sigma^l-p_{+-}\rho^l\\
-\delta p_- \derv{\sigma^l}{x}&=& p_{-,u} \left(\sigma^{l+1}+\sigma^{l-1}-2\sigma^{l}+\delta\derv{\sigma^{l-1}}{x}+\delta\derv{\sigma^{l+1}}{x}\right) +p_{+-}\rho^l-p_{-+}\sigma^l.
\end{eqnarray*}

If $p_{\pm,u}=0$, adding the above two equations gives $p_+\derv{\rho^l}{x}=p_-\derv{\sigma^l}{x}$, meaning that in dilute situation, $\rho$ and $\sigma$ have a linear relationship. This gives a solution to the above ODEs:
\begin{equation}\label{equ_rhoequl0}
\rho(x)=C\exp\left[\left(\frac{p_{-+}N}{p_-}-\frac{p_{+-}N}{p_+}\right)x\right]+D
\end{equation}
\begin{equation}\label{equ_sigmaequl0}
\sigma(x)=\frac{p_+C}{p_-}\exp\left[\left(\frac{p_{-+}N}{p_-}-\frac{p_{+-}N}{p_+}\right)x\right]+\frac{p_{+-}D}{p_{-+}}
\end{equation}
where $C, D$ can be determined by the densities at $\rho(x_0)$ and $\sigma(x_0)$.

If $p_{\pm,u}$ are large ($p_{+-},p_{-+}\ll p_{\pm,u}$), ignoring the turning rates, the above ODEs lead to
\begin{eqnarray}
\delta p_+ \derv{\rho^l}{x} &=& p_{+,u} \left(\rho^{l+1}+\rho^{l-1}-2\rho^{l}-\delta\derv{\rho^{l-1}}{x}-\delta\derv{\rho^{l+1}}{x}\right)\nonumber\\
-\delta p_- \derv{\sigma^l}{x} &=& p_{-,u} \left(\sigma^{l+1}+\sigma^{l-1}-2\sigma^{l}+\delta\derv{\sigma^{l-1}}{x}+\delta\derv{\sigma^{l+1}}{x}\right).\nonumber
\end{eqnarray}

\subsection{Lane-homogeneous densities}
\label{sec_homo}

Now consider homogeneous model which gives lane-homogeneous densities $\rho(x)=\rho^l_i$ due to the symmetric structure of the cylinder and as before $x=i\delta$. In this case we have, ignoring second and higher order terms in $\delta=1/N$, that the stationary state in the dilute case satisfies
\begin{eqnarray}
0 &=& p_{-+}\sigma-p_{+-}\rho+\delta \left[-p_+-p^{l\rightarrow l+1}_{+,u}-p^{l\rightarrow l-1}_{+,u}\right] \derv{\rho}{x}\nonumber\\
0 &=& -p_{-+}\sigma+p_{+-}\rho+\delta \left[p_-+p^{l\rightarrow l+1}_{-,u}+p^{l\rightarrow l-1}_{-,u}\right] \derv{\rho}{x}\nonumber
\end{eqnarray}
Assuming zero net flux, this implies that $v_+\rho-v_-\sigma=0$ where $v_{\pm}/h_s=p_{\pm}+p^{l\rightarrow l+1}_{\pm,u}+p^{l\rightarrow l-1}_{\pm,u}$. Applying the boundary conditions $\rho(0)=\frac{\alpha_+}{p_+M}$ gives
\begin{equation}\label{equ_den_13lane}
\rho(x)=\frac{\alpha_+}{p_+M}\exp\left[\left(\frac{Np_{-+}}{p_{-} + p^{l\rightarrow l-1}_{-,u} + p^{l\rightarrow l+1}_{-,u}}-\frac{Np_{+-}}{p_+ + p^{l\rightarrow l-1}_{+,u} + p^{l\rightarrow l+1}_{+,u}}\right)x\right]
\end{equation}
analogous to the expression found for the two lane model in \cite{Ashwin_etal_2010} but spread over all $M$ lanes. This mean field approximation of density works well for low densities but near the tip, plus-type and minus-type particles do not have complementary density as in \cite{Ashwin_etal_2010}; in fact they can be at the same order near the tip for typical turning rates; for this reason it does not appear to be easy to obtain the mean tip size from a mean field approximation.

\end{document}